\documentclass[journal,10pt]{IEEEtran}
\makeatletter
\def\endthebibliography{%
    \def\@noitemerr{\@latex@warning{Empty `thebibliography' environment}}%
    \endlist
}
\makeatother

\usepackage{cite}

\usepackage[pdftex]{graphicx}
\usepackage[caption=false,font=footnotesize]{subfig}

\usepackage{amsmath} 
\usepackage{mathtools, cuted}
\usepackage{amssymb}
\usepackage{bm}
\usepackage{mathrsfs}
\usepackage{breqn}

\usepackage{pifont}
\usepackage{xcolor}
\usepackage{url}
\usepackage{lettrine}
\usepackage{lipsum}
\usepackage{siunitx}
\usepackage{soul}
\usepackage{array} 
\usepackage[inline]{enumitem}
\usepackage{epsfig}
\usepackage{filecontents}
\usepackage{booktabs}

\usepackage{algpseudocode}
\usepackage{algorithm, tabularx}
\usepackage{multirow}
\newcolumntype{L}[1]{>{\raggedright\let\newline\\\arraybackslash\hspace{0pt}}m{#1}}
\newcolumntype{C}[1]{>{\centering\let\newline\\\arraybackslash\hspace{0pt}}m{#1}}
\newcolumntype{R}[1]{>{\raggedleft\let\newline\\\arraybackslash\hspace{0pt}}m{#1}}
\newlength{\maxwidth}

\makeatletter
\newcommand{\multiline}[1]{%
	\begin{tabularx}{\dimexpr\linewidth-\ALG@thistlm}[t]{@{}X@{}}
		#1
	\end{tabularx}
}
\makeatother
\algdef{SE}[SUBALG]{Indent}{EndIndent}{}{{\algorithmicend\ }}
\algtext*{Indent}
\algtext*{EndIndent}

\usepackage{amsthm}

\theoremstyle{remark}

\begin{document}

\title{Toward Quantum-Enhanced ISAC:\\Active-RIS-Aided Integrated Sensing and Communication with Rydberg Atomic Receivers}

\author{Hong-Bae Jeon,~\IEEEmembership{Member,~IEEE,} Hyung-Joo Moon,~\IEEEmembership{Member,~IEEE,} and Yonghwi Kim,~\IEEEmembership{Member,~IEEE,} 
\thanks{This work was supported by Hankuk University of Foreign Studies Research Fund of 2026. \textit{(Corresponding: Yonghwi Kim.)}}%
\thanks{H.-B. Jeon is with the Department of Information Communications Engineering, Hankuk University of Foreign Studies, Yong-in, 17035, Korea (e-mail: hongbae08@hufs.ac.kr).}%
\thanks{H.-J. Moon is with the School of Integrated Technology, Yonsei University, Seoul, 03722 Korea (e-mail: moonhj@yonsei.ac.kr).}%
\thanks{Y. Kim was with the School of Integrated Technology, Yonsei University, Seoul, 03722 Korea (e-mail: eric\_kim@yonsei.ac.kr).}
}

\maketitle

\begin{abstract}
In this paper, we investigate an active-RIS (ARIS)-aided integrated sensing and communication (ISAC) system with Rydberg Atomic REceiver (RARE). Leveraging the magnitude-only and real-domain observation structure of RARE, we first derive a unified ISAC model, along with a closed-form Cramér-Rao bound (CRB) for direction-of-arrival (DoA) estimation. Based on this formulation, we propose a joint design of the {base station (BS)} beamforming and ARIS reflection coefficients to minimize the CRB under RARE-specific signal-to-interference-noise-ratio (SINR) and ARIS power constraints. To tackle the resulting highly non-convex problem, we develop an alternating optimization (AO) framework that combines semidefinite relaxation (SDR) for beamforming and a majorization-minimization (MM)-based approach for ARIS design. Numerical results demonstrate that the proposed RARE-aware framework significantly outperforms conventional RF-based designs and achieves performance close to the radar-only benchmark, highlighting the potential of RARE for quantum-enhanced ISAC with ARIS.
\end{abstract}


\IEEEpeerreviewmaketitle

\section{Introduction}
\lettrine{I}{ntegrated} sensing and communication (ISAC) has emerged as a key paradigm for sixth-generation (6G) wireless network~\cite{isacjsac, isacmag}, enabling simultaneous data transmission and environmental sensing within a unified framework. By sharing spectrum, hardware, and signaling resources, ISAC improves spectral- and energy-efficiency while supporting emerging applications for 6G such as near-field communications~\cite{hwiisac, nfris}, smart environments~\cite{hjjsac, hjmapxmag, hyoo}, and high-precision localization~\cite{kwonisac, mimorcisac}. In such systems, communication and sensing functionalities are inherently coupled, leading to a fundamental trade-off between communication quality-of-service (QoS) and sensing accuracy. While communication performance is typically characterized by signal-to-interference-plus-noise ratio (SINR) or achievable rate, sensing performance is often quantified by estimation-theoretic metrics such as the Cramér-Rao bound (CRB)~\cite{mimorcisac, liutsp}, which provides a fundamental limit on parameter estimation accuracy.

To enhance ISAC performance, reconfigurable intelligent surfaces (RIS) has been introduced as a promising technology for controllable wireless propagation~\cite{RIST, RISvtm}. By adaptively tuning the phase shifts of incident signals, RIS can reshape the propagation environment, thereby improving wireless coverage and offering passive beamforming gain via programmable elements and its full-duplex nature~\cite{yhRIS, HBRIS}. Building on this capability, RIS-ISAC has recently attracted significant attention, where RIS creates additional controllable propagation paths to simultaneously enhance ISAC performance~\cite{risisacmagg, risisacmag22}. In particular, for sensing, RIS enables coherent combining over the cascaded base station (BS)-RIS-target link, effectively boosting the reflected echo power~\cite{qxisacris}. For communications, RIS enhances the spatial degree-of-freedom (DoF) and coverage by reconfiguring the effective channels~\cite{geisac}.

However, conventional passive-RIS (PRIS) suffers from the multiplicative fading effect, where the cascaded BS-RIS-user/target channel experiences compounded path loss~\cite{DF, vsrelay}, which restricts the achievable performance gains in practical ISAC. This severe attenuation weakens the reflected echo signals for sensing and reduces the increase of effective signal quality for communication, thereby significantly diminishing the potential benefits of RIS in enhancing both sensing and communication performance. To overcome this limitation, active-RIS (ARIS) has been proposed~\cite{aristut, aris1, HBRIS22}, where each reflecting element is equipped with amplification capability in addition to phase control. By compensating for the cascaded signal attenuation through active amplification, ARIS can significantly enhance the effective channel gain and thus improve the overall performance of ISAC~\cite{parisisac, paris22}.

To further unlock the potential of ISAC and overcome the multiplicative fading effect inherent in PRIS, recent studies have actively explored the integration of ARIS into ISAC focusing on sensing-oriented design, communication-centric optimization, and network-level enhancements. From a sensing-centric viewpoint, the authors in~\cite{crbaris} investigated a CRB-minimization in ARIS-ISAC by jointly designing the BS transmit precoding and the ARIS reflection coefficients. Along a similar line,~\cite{arisISACtrx} addressed severe signal attenuation by proposing a joint design of the transmit beamformer, ARIS coefficients, and radar receive filter, aiming to maximize the radar signal power while satisfying user SINR constraints. Beyond sensing performance, security-aware designs have also been investigated. In~\cite{arisISACtvt}, a multi-user ARIS-ISAC in the presence of a malicious unmanned aerial vehicle (UAV) eavesdropper was considered, where ARIS was leveraged to enhance the achievable secrecy rate under radar detection signal-to-noise-ratio (SNR) and transmit power constraints. From a network architecture perspective, in~\cite{arisISACcran}, ARIS was integrated into a cloud radio access network (C-RAN), enabling ISAC from distributed remote radio heads. The design jointly optimized beamforming and fronthaul compression to facilitate target detection in challenging non-line-of-sight (NLoS) environments. Furthermore, mobility-aware and adaptive designs have emerged by exploiting the flexibility of aerial platforms. In~\cite{arisISACdeep}, an aerial-RIS mounted on a UAV was proposed, where deep reinforcement learning was employed to dynamically optimize both the UAV trajectory and beamforming, significantly improving secure energy-efficiency. Similarly,~\cite{arisISACfairness} investigated a UAV-assisted ARIS-ISAC with an emphasis on user fairness, jointly optimizing the BS beamforming, ARIS configuration, and UAV trajectory to maximize the minimum user SINR over time. Overall, these works collectively demonstrate that ARIS provides a powerful and versatile platform for ISAC, thereby substantially mitigating the limitations of conventional PRIS and unlocking new DoF for 6G.

Meanwhile, conventional ISAC primarily rely on classical radio-frequency (RF) receivers, whose performance is fundamentally limited by the Johnson-Nyquist thermal noise (JNTN)~\cite{tnny, quanmobi}. It creates a fundamental sensitivity bottleneck in ISAC under weak-signal conditions, consequently degrading both estimation accuracy and communication performance, even with advanced signal processing techniques. Furthermore, the dependence on antenna-based reception with wavelength-scale structures~\cite{antenna5g, at6g} limits flexible operation across frequency bands, which is particularly detrimental in ISAC that require dynamic spectrum access and multi-band sensing~\cite{TWC24_Cooperative_RIS_ISAC}. As a result, the system becomes less adaptable to heterogeneous environments and diverse sensing scenarios, further exacerbating performance degradation under practical operating conditions.

To overcome this fundamental limitation, Rydberg Atomic Receivers (RARE) have recently emerged as a promising quantum sensing technology~\cite{atomicjsac, atomicmag}. Unlike conventional RF antennas that rely on electronic circuitry for signal reception, RARE exploits the quantum-mechanical response of highly excited Rydberg atoms to incident RF electromagnetic (EM) fields. Specifically, by exciting atoms to high-lying energy states, strong dipole interactions are induced with external RF fields, allowing the incident RF signal to be directly transduced into optical-domain variations via phenomena such as electromagnetically induced transparency (EIT) and Autler-Townes (AT) splitting~\cite{tqe, qprobe}. By leveraging this quantum-optical sensing mechanism, RARE achieves ultra-high field sensitivity with significantly reduced quantum shot noise (QSN)~\cite{qsn}, which is significantly lower than the conventional thermal noise floor~\cite{quanmobi}. This enables near-quantum-limited reception sensitivity on the order of $\mathrm{nVcm^{-1}Hz^{-1/2}}$~\cite{qsens, rarclose3}, thereby overcoming the fundamental noise limitations of classical RF receivers and offering substantial advantages in high-frequency and low-SNR regimes. Moreover, unlike conventional antennas that require physical dimensions comparable to the carrier wavelength for efficient electromagnetic coupling~\cite{antenna5g, at6g}, RARE operates based on photon-atom interactions and does not rely on resonant current induction. This fundamentally different reception mechanism allows RARE to flexibly operate across a wide range of frequencies by selecting appropriate atomic transitions, without requiring hardware reconfiguration~\cite{efm, rydhard1, rydhard2}.

These distinctive properties position RARE as a highly promising technology for next-generation wireless reception, and building upon this potential, recent studies have progressively developed frameworks tailored to RARE. In~\cite{atomicjsac}, atomic demodulation was extended to multi-user settings by casting atomic single/multiple-input-multiple-output (S/MIMO) detection as a biased phase-retrieval problem and proposing algorithm for joint symbol recovery. Subsequently,~\cite{Precoding_atomicMIMO} established an analytic atomic-MIMO model, revealing a fundamental deviation from classical RF MIMO systems due to the nonlinear magnitude-only input-output relation, and introduced in-phase/quadrature (I/Q)-aware precoding strategies to approach atomic-MIMO capacity limits. Furthermore, the architectures such as atomic multi-user uplink detection~\cite{RAR_MU_MIMO_Uplink} and quantum-MIMO receiver arrays~\cite{RAQ_MIMO_Multiband} have demonstrated the feasibility of RARE for spatial multiplexing and multi-user communication, whereas~\cite{saatomic} reveals that spatial variations of atomic quantum states in local-oscillator (LO)-dressed RAREs enable intrinsic beamforming, and develops a segmental vapor-cell architecture that overcomes propagation loss to achieve enhanced beamforming gain and narrower reception patterns.

In parallel, quantum-enhanced spatial sensing techniques have also been investigated. For instance,~\cite{wsat} explored multi-band atomic sensing, demonstrating that RARE can leverage multiple atomic transitions to detect multi-frequency signals and estimate their spatial parameters via quantum probe measurements. In addition,~\cite{TCOM25_Single_RAR_AoA} developed an DoA estimation framework based on a single RARE element by modeling the spatially varying atomic susceptibility induced by interference between the incident and LO fields, leading to a closed-form relation between probe transmission and DoA. Furthermore,~\cite{rydmtdoa} established a RARE-based signal model that explicitly captures LO-induced sensor gain mismatch and, based on this insight, enabled accurate multi-target DoA estimation through a tailored framework.

Complementing these RARE-based spatial sensing approaches,~\cite{cfatomic} proposed a calibration-free reception based on direct/alternating current (D/AC)-Stark response modeling, enabling highly sensitive electric-field detection without external calibration. Collectively, these works establish RARE as a viable and powerful platform, particularly well-suited for ISAC~\cite{kryd}, where reliable detection of weak echo signals and flexible operation across diverse frequency bands are critical. In particular, its ultra-high sensitivity and low-noise characteristics can significantly enhance sensing accuracy~\cite{RAR_Classical_Comm_Sensing, wsat}, while its inherent broadband capability enables more adaptable and efficient ISAC designs. However, despite these promising features, \emph{to the best of our knowledge, a unified system-level framework that \textbf{integrates RARE into propagation-aware ISAC architectures} has not yet been investigated.} Although prior works have investigated RARE-enabled ISAC~\cite{New_Paradigm_RAR_ISAC, isacatomic2}, they primarily focus on receiver design and signal processing aspects, where the propagation environment is assumed to be fixed. Consequently, they cannot leverage propagation-domain DoF to enhance echo signals or suppress multi-user interference, which limits their effectiveness in challenging scenarios and motivates the integration of ARIS for joint propagation and receiver optimization.

Herein, the complementary gains can be achieved from both the propagation and receiver domains. Specifically, ARIS actively reshapes and amplifies the channel to mitigate severe path loss and multiplicative fading, while RARE enables high-sensitivity signal acquisition with a reduced noise floor. This complementary interaction establishes a powerful synergy, jointly enabling reliable ISAC beyond the capability of conventional RF-based systems. Motivated by this, this paper investigates an ARIS-ISAC framework with RARE that explicitly incorporates the RARE-specific signal model into a unified ISAC framework. This reveals new design insights arising from its real-domain observation structure and low-noise characteristics of RARE, which fundamentally reshape the signal model and require a rethinking of joint communication and sensing design. The main contributions of this paper are summarized as follows:

\begin{itemize}
\item We establish a unified ARIS-ISAC framework incorporating RARE at multiple communication users and the sensing receiver at the BS. This dual deployment simultaneously mitigates receiver-side noise limitations and improves echo signal acquisition, thereby enhancing both communication reliability and sensing accuracy. Moreover, we characterize the unique magnitude-only real-domain observation structure and show that the RARE measurement can be equivalently represented as a structured real-valued projection of the effective channel, revealing that communication information is intrinsically embedded in the power-domain observations of RARE.

\item Based on the proposed RARE-aware signal model, we derive a closed-form CRB for target DoA estimation in an ARIS-ISAC with RARE at BS. It explicitly captures the impact of ARIS-induced channel shaping, amplified noise propagation, and RARE-specific measurement characteristics, thereby providing key insights into how receiver sensitivity and propagation control jointly influence sensing accuracy.

\item Leveraging the above analysis, we formulate a CRB-driven joint design problem, where the transmit beamforming and ARIS coefficients are jointly optimized under RARE-specific SINR and ARIS power constraints. To efficiently solve the resulting highly non-convex problem, we develop a structured alternating optimization (AO) framework:
\begin{enumerate}
\item We first reformulate the transmit beamforming subproblem into a tractable semidefinite relaxation (SDR) by lifting the beamforming vector into a higher-dimensional positive semidefinite matrix. This converts the original non-convex quadratic problem into a convex semidefinite program (SDP), enabling efficient optimization under the QoS constraints. The beamforming solution is then recovered through rank-one approximation, yielding a near-optimal transmit strategy with manageable complexity.
\item We then develop an ARIS update strategy based on the majorization-minimization (MM) framework to handle the non-convex unit-modulus and power-coupled constraints introduced by ARIS amplification. Specifically, a tight surrogate function is constructed to upper-bound the original objective while capturing both desired signal enhancement and amplified noise propagation. The resulting subproblem admits a low-complexity element-wise update with guaranteed monotonic improvement across iterations.
\end{enumerate}

\item Extensive simulation demonstrates that the proposed RARE-enabled ARIS-ISAC framework consistently outperforms conventional RF-based designs across various operating conditions and approaches the radar-only performance bound. The results highlight the effectiveness of jointly exploiting ARIS-driven propagation enhancement and RARE-based quantum-enhanced reception, enabling robust echo recovery for accurate sensing while simultaneously improving communication reliability, thereby effectively enhancing overall ISAC performance.
\end{itemize}

\begin{figure}[t]
	\begin{center}
		\includegraphics[width=0.97\columnwidth,keepaspectratio]%
		{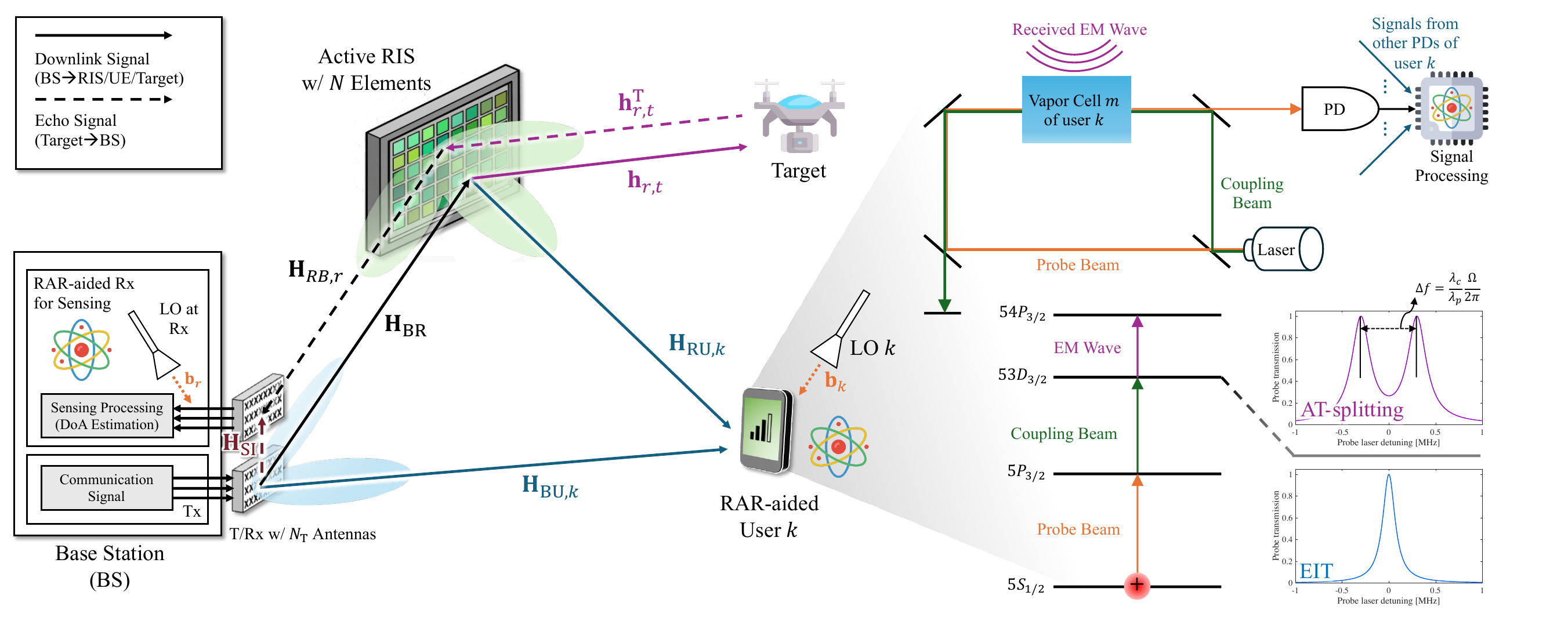}
		\caption{(Left) System model of ARIS-ISAC with RARE at BS Rx and users and (Right) illustration of signal processing in RAR, where the resulting spectral separation $\Delta f$ of AT-splitting phenomenon is subsequently mapped to the corresponding Rabi frequency $\Omega$.}
		\label{fig_sys}
	\end{center}
\end{figure}
\section{System Model}
\subsection{ARIS-RARE Communication Signal Model}
As illustrated in the left-side of Fig.~\ref{fig_sys}, we consider an ARIS-ISAC system where a dual-functional BS simultaneously supports downlink multi-user communication and radar sensing. The BS is equipped with $N_t$ Tx antennas for ISAC signal transmission and $N_r$-cell Rx RARE for echo acquisition for target estimation, with $N_r = N_t$ for simplicity~\cite{crbaris}. The system further comprises an $N$-element ARIS, $K$ downlink users each equipped with an $M$-cell RARE, and a single point target to be sensed. The BS employs a transmit waveform $
\mathbf{s}=[s_1 \cdots s_K]^{\mathrm{T}} \in \mathbb{C}^{K}$ normalized as $\mathbb{E}[\mathbf{s}\mathbf{s}^*] = \mathbf{I}_K$ to concurrently deliver symbols to $K$ users and probe the environment, thereby enabling simultaneous sensing and communication, particularly for target detection within the BS blind zone. The BS employs a linear precoder $\mathbf{W}\in\mathbb{C}^{N_t\times K}$ to spatially multiplex the user symbols, generating the transmit signal $\mathbf{x} = \mathbf{W}\mathbf{s} \in \mathbb{C}^{N_t}$. Accordingly, the transmit covariance matrix becomes $\mathbf{R}_x =\mathbb{E}[\mathbf{x}\mathbf{x}^*]=\mathbf{W}\mathbf{W}^*$, and the transmit power at BS is constrained by $\mathrm{tr}(\mathbf{W}\mathbf{W}^*) \le P_{\max}$. Let $\mathbf{H}_{BR}\in\mathbb{C}^{N\times N_t}$ denote the BS-ARIS channel that characterizes the propagation between the BS antenna array and the ARIS elements, and let $\mathbf{H}_{RU,k} \in \mathbb{C}^{M\times N}$ and $\mathbf{H}_{BU,k} \in \mathbb{C}^{M\times N_t}$ denote the ARIS-RARE channel and the direct BS-RARE channel associated with the $k$th user, respectively. Each RARE user consists of $M$ atomic sensing cells, which measure the incident RF field.

The atomic channels are modeled using a multipath Rydberg EM coupling framework~\cite{Precoding_atomicMIMO}, which captures the physical interaction between incident multipath EM waves and Rydberg atomic transitions. Specifically, the $(m,n)$th element of $\mathbf H_{\mathrm{RU},k}$ is expressed as~\cite{atomicjsac}
\begin{equation}
[\mathbf H_{\mathrm{RU},k}]_{m,n}=
\sum_{\ell=1}^{L_0}
\frac{1}{\hbar}
\boldsymbol\mu_{\mathrm{eg}}^{\mathrm T}
\boldsymbol\epsilon_{m,n,k,\ell}
\rho_{m,n,k,\ell}
e^{j\phi_{m,n,k,\ell}},
\label{eq:H_RU_elem}
\end{equation}
where $L_0$ denotes the number of propagation paths, and $\hbar$ is the reduced Planck constant. Here, $\boldsymbol\mu_{\mathrm{eg}}\in\mathbb R^{3}$ represents the electric dipole moment associated with the Rydberg transition, while $\boldsymbol\epsilon_{m,n,k,\ell}\in\mathbb R^{3}$, $\rho_{m,n,k,\ell}$, and $\phi_{m,n,k,\ell}$ correspond to the polarization vector, path attenuation, and phase of the $\ell$th multipath component linking ARIS element $n$ to the $m$th vapor cell of user $k$, respectively. An analogous formulation applies to ${\mathbf H_{\mathrm{BU},k}}$ with the corresponding propagation parameters.

For ARIS, let $\boldsymbol{\Phi} = \mathrm{diag}(\phi_1,\cdots,\phi_N)$ denote the reflection matrix of the ARIS, where the reflection coefficient of the $n$th ARIS element is modeled as $\phi_n = a_n e^{j\theta_n}~( a_n\in(0,  a_{\max}],~\theta_n\in[0, 2\pi))$. Herein, $a_n$ represents the amplification factor and $\theta_n$ denotes the controllable phase shift. The parameter $a_{\max}$ specifies the maximum amplification gain allowed by the ARIS hardware.

The RF signal observed at the input of the $k$th RARE is therefore given by
\begin{equation}
\mathbf{r}_k=\mathbf{H}_{BU,k}\mathbf{x}+\mathbf{H}_{RU,k}\boldsymbol{\Phi}\left(\mathbf{H}_{BR}\mathbf{x}+\mathbf{z}_0\right)+\mathbf{b}_k+\mathbf{n}_k,
\label{eq:RAR_RF}
\end{equation}
where $\mathbf b_k\in\mathbb C^{M}$ denotes the LO RF field associated with user $k$, and $\mathbf n_k\sim\mathcal{CN}(\mathbf 0,\sigma_{q,k}^2\mathbf I_M)$ represents the quantum shot noise (QSN) from RARE, which is several orders of magnitude lower than the conventional JNTN~\cite{quanmobi}. The $m$th element of the LO signal, $b_{k,m}$, is modeled as
\begin{equation}
  b_{k,m}  =  \frac{s_b}{\hbar}
  \boldsymbol\mu_{\mathrm{eg}}^{\mathrm T}
  \boldsymbol\epsilon_{k,m,b}
  \rho_{k,m,b}
  \sqrt{P_b}e^{j\phi_{k,m,b}}
  =
  |b_{k,m}|e^{j\angle b_{k,m}},
  \label{eq:b_k_m}
\end{equation}
where $s_b$ is a known reference symbol, $P_b$ denotes the LO power, and $\boldsymbol\epsilon_{k,m,b}$, $\rho_{k,m,b}$, and $\phi_{k,m,b}$ correspond to the polarization vector, path attenuation, and phase of the LO signal, respectively~\cite{Precoding_atomicMIMO, atomicjsac}. In addition, $\mathbf{z}_0$ models the amplification noise at the ARIS and is given by $\mathbf{z}_0 \sim \mathcal{CN}(\mathbf{0}, \sigma_z^2 \mathbf{I}_N)$, where $\sigma_z^2$ denotes the amplifier noise power.

Hence, by defining the effective BS-RARE channel as $\mathbf{H}_{\mathrm{eff},k}(\boldsymbol{\Phi})
=
\mathbf{H}_{BU,k}
+
\mathbf{H}_{RU,k}\boldsymbol{\Phi}\mathbf{H}_{BR}$, and~\eqref{eq:RAR_RF} becomes
\begin{equation}
\mathbf{r}_k
=
\sum_{i=1}^{K}
\mathbf{H}_{\mathrm{eff},k}(\boldsymbol{\Phi})\mathbf{w}_i s_i
+
\mathbf{H}_{RU,k}\boldsymbol{\Phi}\mathbf{z}_0
+
\mathbf{b}_k+\mathbf{n}_k,
\label{eq:RAR_RF3}
\end{equation}
where $\mathbf{w}_i$ denotes the $i$th column of $\mathbf{W}$. Herein, the first term captures the desired signal along with multi-user interference, whereas the remaining terms account for the contributions of ARIS amplification noise, LO signal, and measurement noise.

\subsection{Optical Readout of RARE in Communication-Side}
\label{orrc}
As illustrated in the right-side of Fig.~\ref{fig_sys}, the incident RF signal interacts with the Rydberg atomic transitions, inducing EIT and AT splitting~\cite{efm}. The resulting AT-splitting interval $\{\Delta f_m\}$ is directly related to the corresponding Rabi frequency $\{\Omega_m\}$~\cite{Precoding_atomicMIMO, fox}, which can be expressed as $\Delta f_m=\frac{\lambda_c}{\lambda_p}\frac{\Omega_m}{2\pi}~(\forall m)$, where $\lambda_c$ and $\lambda_p$ represent the wavelengths of the coupling and probe beams, respectively. Accordingly, by measuring $\Delta f_m$, $\Omega_m$ can be obtained, which in turn enables the recovery of the transmitted signal since $\{\Omega_m\}$ equals to the magnitude of $\mathbf r_k$~\cite{atomicjsac}. Therefore, the resulting optical measurement can be expressed as $\tilde{\mathbf{z}}_k=\left|\mathbf{r}_k \right|$, where the absolute value is applied element-wise, and each real-valued component represents the corresponding $\{\Omega_m\}$.

When the LO field dominates the received RF field (strong-reference regime), the nonlinear magnitude measurement can be linearized via a first-order expansion around $\mathbf b_k$~\cite{Precoding_atomicMIMO}:
\begin{equation}
{\mathbf{z}}_k\triangleq \tilde{\mathbf{z}}_k-|\mathbf b_k|
\approx
\Re
\left\{
\mathbf{D}_{b,k}
\mathbf{H}_{\mathrm{eff},k}(\boldsymbol{\Phi})
\mathbf x
\right\}
+
\boldsymbol{\nu}_k
\label{eq:RAR_real_model2}
\end{equation}
where $\boldsymbol{\nu}_k
=
\Re
\left\{
\mathbf{D}_{b,k}
\mathbf{H}_{RU,k}\boldsymbol{\Phi}\mathbf z_0
\right\}
+
\boldsymbol{\eta}_k$ with $\boldsymbol{\eta}_k\sim\mathcal N\left(\mathbf 0,\frac{\sigma_{q,k}^2}{2}\mathbf I_{M}\right)$, denotes the effective noise consisting of the forwarded ARIS noise and the optical readout noise.

Stacking $\{\mathbf z_k\}$ yields the global communication observation vector: $\mathbf{z}_{\mathrm{com}}
=
\begin{bmatrix}
{\mathbf{z}}_1\\
\vdots\\
{\mathbf{z}}_K
\end{bmatrix}
\in\mathbb{R}^{KM}$. By defining the effective communication channel matrix $\mathbf{E}_{\mathrm{com}}(\boldsymbol{\Phi})
=
\begin{bmatrix}
\mathbf{D}_{b,1}\mathbf{H}_{\mathrm{eff},1}(\boldsymbol{\Phi})\\
\vdots\\
\mathbf{D}_{b,K}\mathbf{H}_{\mathrm{eff},K}(\boldsymbol{\Phi})
\end{bmatrix}
\in\mathbb{C}^{KM\times N_t}$, the stacked communication model can therefore be compactly written as
\begin{equation}
\mathbf{z}_{\mathrm{com}}
=
\Re
\left\{
\mathbf{E}_{\mathrm{com}}(\boldsymbol{\Phi})
\mathbf{x}
\right\}
+
\boldsymbol{\nu}_{\mathrm{com}}
\label{eq:stacked_model}
\end{equation}
where $\boldsymbol{\nu}_{\mathrm{com}}
=
\begin{bmatrix}
\boldsymbol{\nu}_1\\
\vdots\\
\boldsymbol{\nu}_K
\end{bmatrix}$. 

\subsection{Radar Echo Model and RARE-Based Sensing Observation}
\label{subsec:radar_rare_model}
\subsubsection{Echo Signal Model with RARE}
We consider ARIS-assisted sensing at the BS, where the target echo is received through an $N$-element ARIS. In practice, despite advanced TRx isolation and cancellation, a residual self-interference (SI) component generally remains at the sensing receiver~\cite{crbaris, fdisac}. Accordingly, the received RF echo field at the sensing-side RARE in the $\ell$th snapshot is modeled as
\begin{equation}
\begin{aligned}
\mathbf{r}_{r}[\ell]
=&
\alpha 
\mathbf{H}_{RB,r}^{\mathrm{T}}
\boldsymbol{\Phi}
\mathbf{h}_{r,t}(\theta)
\mathbf{h}_{r,t}^{\mathrm{T}}(\theta)
\boldsymbol{\Phi}
\mathbf{H}_{BR}
\mathbf{x}[\ell]\\
&+\mathbf{H}_{\mathrm{SI}}\mathbf{x}[\ell]+\mathbf{H}_{RB,r}^{\mathrm{T}}\boldsymbol{\Phi}\mathbf{z}_1[\ell]+\mathbf b_r [\ell]+\mathbf{n}_r[\ell],
\end{aligned}
\label{eq:radar_rf_echo_rare_single}
\end{equation}
where $\alpha\in\mathbb{C}$ denotes the complex target reflection coefficient, $\theta$ is the target DoA with respect to the ARIS, $\mathbf{h}_{r,t}(\theta)$ is the direct link between ARIS-target, which is assumed to be line-of-sight (LoS) link~\cite{crbaris, Liu_RIS_ISAC_CRB}, $\mathbf{H}_{RB,r}^{\mathrm{T}}\in\mathbb C^{N_t\times N}$ is the ARIS-RARE (BS) channel denoted by its $(m,n)$th element:
\begin{equation}
[\mathbf{H}_{RB,r}^{\mathrm{T}}]_{m,n}=
\sum_{\ell=1}^{L_r}
\frac{1}{\hbar}
\boldsymbol\mu_{\mathrm{eg}}^{\mathrm T}
\boldsymbol\epsilon_{m,n,r,\ell}
\rho_{m,n,r,\ell}
e^{j\phi_{m,n,r,\ell}},
\label{eq:H_RB_elem}
\end{equation}
where the related parameters are similarly defined as before. $\mathbf{H}_{\mathrm{SI}}\in\mathbb{C}^{N_t\times N_t}$ denotes the residual SI channel after cancellation~\cite{fdisac}. Moreover, $\mathbf b_r [\ell]$ is the LO field for BS echo, which follows the same structure in~\eqref{eq:b_k_m} with corresponding parameters, $\mathbf{z}_1[\ell]\sim\mathcal{CN}(\mathbf{0},\sigma_z^2\mathbf{I}_{N})$ represents the ARIS-side amplifier noise, and $\mathbf{n}_r[\ell]\sim\mathcal{CN}(\mathbf{0},\sigma_{q,r}^2\mathbf{I}_{N_t})$ is the QSN associated with the sensing RF field received with RARE. By defining $\mathbf{Q}(\boldsymbol{\Phi},\theta)
\triangleq
\mathbf{H}_{RB,r}^{\mathrm{T}}
\boldsymbol{\Phi}
\mathbf{h}_{r,t}(\theta)
\mathbf{h}_{r,t}^{\mathrm{T}}(\theta)
\boldsymbol{\Phi}
\mathbf{H}_{BR}
\in\mathbb{C}^{N_t\times N_t}$,~\eqref{eq:radar_rf_echo_rare_single} can be compactly rewritten as
\begin{equation}
\mathbf{r}_{r}[\ell]
=
\alpha \mathbf{Q}(\boldsymbol{\Phi},\theta)\mathbf{x}[\ell]
+
\mathbf{H}_{\mathrm{SI}}\mathbf{x}[\ell]
+
\mathbf{H}_{RB,r}^{\mathrm{T}}\boldsymbol{\Phi}\mathbf{z}_1[\ell]
+
\mathbf b_r[\ell]+\mathbf{n}_r[\ell].
\label{eq:radar_rf_echo_rare_compact}
\end{equation}

\subsubsection{RARE Optical Readout Model for Sensing}
Consistent with Section~\ref{orrc}, the raw optical measurement of sensing signal at the $\ell$th snapshot is expressed as $
\tilde{\mathbf{z}}_r[\ell]
=
\left|
\mathbf{r}_{r}[\ell]
\right|$. Under the strong-reference regime~\cite{Precoding_atomicMIMO}, by subtracting $|\mathbf b_r[\ell]|$, the sensing observation can be compactly written as
\begin{equation}
\mathbf z_r[\ell]\triangleq \tilde{\mathbf z}_r[\ell]-|\mathbf b_r[\ell]|
\approx
\Re
\left\{
\mathbf E_r(\boldsymbol{\Phi},\theta)
\mathbf x[\ell]
\right\}
+
\mathbf w_r[\ell],
\label{eq:radar_real_model_compact}
\end{equation}
where the effective sensing channel is defined as
\begin{equation}
\mathbf E_r(\boldsymbol{\Phi},\theta)
\triangleq
\mathbf D_{b,r}
\left(
\alpha \mathbf Q(\boldsymbol{\Phi},\theta)
+
\mathbf H_{\mathrm{SI}}
\right)
\label{eq:Er_def}
\end{equation}
with $\mathbf{D}_{b,r}
\triangleq
\mathrm{diag}
\left(
e^{-j\angle b_{r,1}},
\cdots,
e^{-j\angle b_{r,N_t}}
\right)$, and the effective noise is given by $\mathbf w_r[\ell]
=
\Re
\left\{
\mathbf D_{b,r}
\mathbf{H}_{RB,r}^{\mathrm{T}}\boldsymbol{\Phi}\mathbf z_1[\ell]
\right\}
+
\boldsymbol{\eta}_r[\ell]$, where $\boldsymbol{\eta}_r[\ell]\sim\mathcal N\left(\mathbf 0,\frac{\sigma_{q,r}^2}{2}\mathbf I_{N_t}\right)$ denotes the real-valued readout noise induced by the atomic measurement process at BS. Since $\mathbf{z}_1[\ell]\sim\mathcal{CN}(\mathbf{0},\sigma_z^2\mathbf{I}_{N})$ and $\mathbf{D}_{b,r}$ is diagonal unitary, the projected noise $\Re
\left\{
\mathbf D_{b,r}
\mathbf{H}_{RB,r}^{\mathrm{T}}\boldsymbol{\Phi}\mathbf z_1[\ell]
\right\}
$ and hence $\mathbf w_r[\ell]$ remain real Gaussian: $\mathbf{w}_r[\ell]\sim \mathcal N(\mathbf 0,\mathbf R_{w})$ with
\begin{equation}
\mathbf R_w
=
\frac{1}{2}
\sigma_z^2\Re\{
\mathbf{D}_{b,r}
\mathbf{H}_{RB,r}^{\mathrm{T}}
\boldsymbol{\Phi}\boldsymbol{\Phi}^{*}
\bar{\mathbf{H}}_{RB,r}
\mathbf{D}_{b,r}^{*}\}
+
\frac{\sigma_{q,r}^2}{2}\mathbf I_{N_t},
\label{eq:Rw_def}
\end{equation}
where $\bar{\mathbf{H}}_{RB,r}\triangleq\mathbf{H}_{RB,r}^{\mathrm{T}*}$ (i.e., element-wise conjugate).

Stacking $L$ snapshots, define $\mathbf Z_r
\triangleq
[\mathbf z_r[1] \cdots \mathbf z_r[L]]
\in\mathbb R^{N_t\times L}$. Then $\mathbf Z_r=\Re\left\{\mathbf E_r(\boldsymbol\Phi,\theta)\mathbf X\right\}+\mathbf W_r$, where $\mathbf X=[\mathbf x[1] \cdots \mathbf x[L]]$ and $\mathbf W_r=[\mathbf w_r[1] \cdots \mathbf w_r[L]]$. Finally, vectorizing it, we obtain the real-valued sensing observation model
\begin{equation}
\mathbf z_e
\triangleq
\mathrm{vec}(\mathbf Z_r)
=
\boldsymbol{\mu}(\boldsymbol\xi)
+
\mathbf w,
\label{eq:real_obs_model}
\end{equation}
where $\boldsymbol{\mu}(\boldsymbol\xi)
\triangleq
\Re\left\{
\alpha \mathbf a(\theta)+\mathbf b
\right\}
\in\mathbb R^{N_tL}$ with $\mathbf a(\theta)
\triangleq
\mathrm{vec}
\Big(
\mathbf D_{b,r}\mathbf Q(\boldsymbol\Phi,\theta)\mathbf X
\Big)$ and $\mathbf b
\triangleq
\mathrm{vec}
\Big(
\mathbf D_{b,r}\mathbf H_{\mathrm{SI}}\mathbf X
\Big)$, and $\mathbf w\sim\mathcal N(\mathbf 0,\mathbf R_e),
~
\mathbf R_e
\triangleq
\mathbf I_L\otimes \mathbf R_w$ with Kronecker product $\otimes$.

\subsection{CRB for DoA Estimation}
\label{subsec:crb_rare_sensing}

We define the real-valued unknown parameter vector as $\boldsymbol\xi
\triangleq
[\theta~\alpha_R~\alpha_I]^{\mathrm T}\in\mathbb R^3$, where $\alpha_R\triangleq \Re\{\alpha\}$ and $\alpha_I\triangleq \Im\{\alpha\}$. 
Let $\mathbf S=[\mathbf s[1] \cdots \mathbf s[L]]\in\mathbb C^{K\times L}$ denote the probing symbol matrix, where $\mathbf X=\mathbf W\mathbf S$. Since the BS is dual-functional, $\mathbf X$ is perfectly known at the BS, and $L$ is sufficiently large for estimation, $\mathbf S$ is designed to satisfy $\mathbf S\mathbf S^{*}=L\mathbf I_K$~\cite{Liu_RIS_ISAC_CRB, crbaris}:
\begin{equation}
\mathbf X\mathbf X^{*}
=
\mathbf W\mathbf S\mathbf S^{*}\mathbf W^{*}
=
L\mathbf W\mathbf W^{*}.
\label{eq:XXH_rare}
\end{equation}

Define $\dot{\mathbf Q}(\boldsymbol\Phi,\theta)
\triangleq
\frac{\partial \mathbf Q(\boldsymbol\Phi,\theta)}{\partial \theta}$ and $\dot{\mathbf a}(\theta)
\triangleq
\frac{\partial \mathbf a(\theta)}{\partial \theta}
=
\mathrm{vec}
\Big(
\mathbf D_{b,r}\dot{\mathbf Q}(\boldsymbol\Phi,\theta)\mathbf X
\Big)$. Since $\boldsymbol\mu(\boldsymbol\xi)
=
\Re\{\alpha \mathbf a(\theta)+\mathbf b\}$, the partial derivatives are given by
\begin{equation}
\frac{\partial \boldsymbol\mu}{\partial \theta}
=
\Re\{\alpha \dot{\mathbf a}(\theta)\},
~
\frac{\partial \boldsymbol\mu}{\partial \alpha_R}
=
\Re\{\mathbf a(\theta)\},
~
\frac{\partial \boldsymbol\mu}{\partial \alpha_I}
=
\Re\{j\mathbf a(\theta)\}.
\label{eq:dmu_rare}
\end{equation}
Noting that $\Re\{j\mathbf a\}=-\Im\{\mathbf a\}$, the above derivatives are all real-vectors in $\mathbb R^{N_tL}$, as required by the real Gaussian model.

With~\eqref{eq:real_obs_model}, the $(i,j)$th entry of the Fisher information matrix (FIM) $\mathbf M$ is: $\mathbf M_{ij}
=
\left(
\frac{\partial \boldsymbol\mu}{\partial \xi_i}
\right)^{\mathrm T}
\mathbf R_e^{-1}
\left(\frac{\partial \boldsymbol\mu}{\partial \xi_j}
\right)
+
\frac{1}{2}
\mathrm{tr}
\left\{
\mathbf R_e^{-1}
\frac{\partial \mathbf R_e}{\partial \xi_i}
\mathbf R_e^{-1}
\frac{\partial \mathbf R_e}{\partial \xi_j}
\right\}$. Herein, since $\mathbf R_e\triangleq\mathbf I_L\otimes \mathbf R_w$ is independent of $\boldsymbol\xi$, we have $\frac{\partial \mathbf R_e}{\partial \xi_i}=\mathbf 0~ (\forall i)$, and hence
\begin{equation}
\mathbf M_{ij}
=
\left(
\frac{\partial \boldsymbol\mu}{\partial \xi_i}
\right)^{\mathrm T}
\mathbf R_e^{-1}
\left(
\frac{\partial \boldsymbol\mu}{\partial \xi_j}
\right).
\label{eq:FIM_real_simplified}
\end{equation}
For compactness, define $\mathbf u_\theta
\triangleq
\Re\{\alpha \dot{\mathbf a}\},
~
\mathbf u_R
\triangleq
\Re\{\mathbf a\},
~
\mathbf u_I
\triangleq
\Re\{j\mathbf a\}
=
-\Im\{\mathbf a\}$. Then the FIM can be written as
\begin{equation}
\mathbf M
=
\begin{bmatrix}
\mathbf u_\theta^{\mathrm T}\mathbf R_e^{-1}\mathbf u_\theta
&
\mathbf u_\theta^{\mathrm T}\mathbf R_e^{-1}\mathbf u_R
&
\mathbf u_\theta^{\mathrm T}\mathbf R_e^{-1}\mathbf u_I
\\
\mathbf u_R^{\mathrm T}\mathbf R_e^{-1}\mathbf u_\theta
&
\mathbf u_R^{\mathrm T}\mathbf R_e^{-1}\mathbf u_R
&
\mathbf u_R^{\mathrm T}\mathbf R_e^{-1}\mathbf u_I
\\
\mathbf u_I^{\mathrm T}\mathbf R_e^{-1}\mathbf u_\theta
&
\mathbf u_I^{\mathrm T}\mathbf R_e^{-1}\mathbf u_R
&
\mathbf u_I^{\mathrm T}\mathbf R_e^{-1}\mathbf u_I
\end{bmatrix}.
\label{eq:FIM_real_matrix}
\end{equation}
Partition the FIM as $\mathbf M
=
\begin{bmatrix}
M_{\theta\theta} & \mathbf M_{\theta\alpha}\\
\mathbf M_{\theta\alpha}^{\mathrm T} & \mathbf M_{\alpha\alpha}
\end{bmatrix},
~
\boldsymbol\alpha\triangleq[\alpha_R~\alpha_I]^{\mathrm T}$, where $M_{\theta\theta}
=
\mathbf u_\theta^{\mathrm T}\mathbf R_e^{-1}\mathbf u_\theta,~\mathbf M_{\theta\alpha}
=
\begin{bmatrix}
\mathbf u_\theta^{\mathrm T}\mathbf R_e^{-1}\mathbf u_R
&
\mathbf u_\theta^{\mathrm T}\mathbf R_e^{-1}\mathbf u_I
\end{bmatrix},~\mathbf M_{\alpha\alpha}
=
\begin{bmatrix}
\mathbf u_R^{\mathrm T}\mathbf R_e^{-1}\mathbf u_R
&
\mathbf u_R^{\mathrm T}\mathbf R_e^{-1}\mathbf u_I
\\
\mathbf u_I^{\mathrm T}\mathbf R_e^{-1}\mathbf u_R
&
\mathbf u_I^{\mathrm T}\mathbf R_e^{-1}\mathbf u_I
\end{bmatrix}$. Therefore, the CRB of $\theta$ is given by the Schur complement
\begin{equation}
\mathrm{CRB}_{\theta}(\mathbf W,\boldsymbol\Phi)
=
\left(
M_{\theta\theta}
-
\mathbf M_{\theta\alpha}
\mathbf M_{\alpha\alpha}^{-1}
\mathbf M_{\theta\alpha}^{\mathrm T}
\right)^{-1}.
\label{eq:CRB_theta_rare_general_re}
\end{equation}
For convenience, define
\begin{equation}
\begin{aligned}
&A
\triangleq
\mathbf u_R^{\mathrm T}\mathbf R_e^{-1}\mathbf u_R,
~
B
\triangleq
\mathbf u_I^{\mathrm T}\mathbf R_e^{-1}\mathbf u_I,
~
C
\triangleq
\mathbf u_R^{\mathrm T}\mathbf R_e^{-1}\mathbf u_I,\\
&p
\triangleq
\mathbf u_\theta^{\mathrm T}\mathbf R_e^{-1}\mathbf u_R,
~
q
\triangleq
\mathbf u_\theta^{\mathrm T}\mathbf R_e^{-1}\mathbf u_I,
~
r
\triangleq
\mathbf u_\theta^{\mathrm T}\mathbf R_e^{-1}\mathbf u_\theta.
\label{eq:ABC_def_re}
\end{aligned}
\end{equation}
Then $\mathbf M_{\alpha\alpha}=
\begin{bmatrix}
A & C\\
C & B
\end{bmatrix}$ and $\mathbf M_{\theta\alpha}
=[p~q]$, and by substituting into~\eqref{eq:CRB_theta_rare_general_re}, we obtain
\begin{equation}
\mathrm{CRB}_{\theta}(\mathbf W,\boldsymbol\Phi)
=
\left(
r-
\frac{Bp^2-2Cpq+Aq^2}{AB-C^2}
\right)^{-1}.
\label{eq:CRB_theta_rare_closed_re}
\end{equation}
To further express~\eqref{eq:CRB_theta_rare_closed_re} in terms of $\mathbf a$ and $\dot{\mathbf a}$, define
\begin{equation}
\begin{aligned}
&u
\triangleq
\mathbf a^{*}\mathbf R_e^{-1}\mathbf a,
~
\tilde u
\triangleq
\mathbf a^{\mathrm T}\mathbf R_e^{-1}\mathbf a,~t
\triangleq
\dot{\mathbf a}^{*}\mathbf R_e^{-1}\mathbf a,\\
&\tilde t
\triangleq
\dot{\mathbf a}^{\mathrm T}\mathbf R_e^{-1}\mathbf a,~v
\triangleq
\dot{\mathbf a}^{*}\mathbf R_e^{-1}\dot{\mathbf a},
~
\tilde v
\triangleq
\dot{\mathbf a}^{\mathrm T}\mathbf R_e^{-1}\dot{\mathbf a}.
\label{utvt}
\end{aligned}
\end{equation}
By using $\mathbf u_R
=
\Re\{\mathbf a\}
=
\frac{\mathbf a+\mathbf a^{*}}{2},
~
\mathbf u_I
=
\Re\{j\mathbf a\}
=
-\Im\{\mathbf a\}
=
\frac{j(\mathbf a-\mathbf a^{*})}{2},~\mathbf u_\theta
=
\Re\{\alpha\dot{\mathbf a}\}
=
\frac{\alpha\dot{\mathbf a}+\alpha^{*}\dot{\mathbf a}^{*}}{2}$,~\eqref{eq:ABC_def_re} becomes
\begin{equation}
\begin{aligned}
\label{abcpqr}
&A=\frac{u+\Re\{\tilde u\}}{2},
~
B=\frac{u-\Re\{\tilde u\}}{2},
~
C=-\frac{\Im\{\tilde u\}}{2},\\
&p=\frac{\Re\{\alpha^{*}t\}+\Re\{\alpha\tilde t\}}{2},
~
q=-\frac{\Im\{\alpha^{*}t\}+\Im\{\alpha\tilde t\}}{2},\\
&r=\frac{|\alpha|^2 v+\Re\{\alpha^2\tilde v\}}{2}.
\end{aligned}
\end{equation}
In addition, using $\mathrm{vec}(\mathbf A)^{*}
(\mathbf I_L\otimes \mathbf B)
\mathrm{vec}(\mathbf C)
=
\mathrm{tr}\{\mathbf A^{*}\mathbf B\mathbf C\}$, together with $\mathbf R_e=\mathbf I_L\otimes \mathbf R_w$ and $\mathbf S\mathbf S^*=L\mathbf I_K$, we have
\begin{equation}
\begin{aligned}
u
&=
\mathrm{tr}
\left\{
\mathbf X^{*}
\mathbf Q^{*}
\mathbf D_{b,r}^{*}
\mathbf R_w^{-1}
\mathbf D_{b,r}
\mathbf Q
\mathbf X
\right\}\\
&=
L \mathrm{tr}
\left\{
\mathbf Q
\mathbf W\mathbf W^{*}
\mathbf Q^{*}
\mathbf D_{b,r}^{*}
\mathbf R_w^{-1}
\mathbf D_{b,r}
\right\},\\
t
&=
\mathrm{tr}
\left\{
\mathbf X^{*}
\dot{\mathbf Q}^{*}
\mathbf D_{b,r}^{*}
\mathbf R_w^{-1}
\mathbf D_{b,r}
\mathbf Q
\mathbf X
\right\}\\
&
=
L \mathrm{tr}
\left\{
\dot{\mathbf Q}
\mathbf W\mathbf W^{*}
\mathbf Q^{*}
\mathbf D_{b,r}^{*}
\mathbf R_w^{-1}
\mathbf D_{b,r}
\right\},\\
v
&=
\mathrm{tr}
\left\{
\mathbf X^{*}
\dot{\mathbf Q}^{*}
\mathbf D_{b,r}^{*}
\mathbf R_w^{-1}
\mathbf D_{b,r}
\dot{\mathbf Q}
\mathbf X
\right\}\\
&=
L \mathrm{tr}
\left\{
\dot{\mathbf Q}
\mathbf W\mathbf W^{*}
\dot{\mathbf Q}^{*}
\mathbf D_{b,r}^{*}
\mathbf R_w^{-1}
\mathbf D_{b,r}
\right\},
\label{utvtrace}
\end{aligned}
\end{equation}
and since $\{s_i\}$ are unit-power proper signal: $\mathbb{E}[\mathbf S\mathbf S^{\mathrm T}]
=
\sum_{\ell=1}^{L}\mathbb{E}[\mathbf s[\ell]\mathbf s^{\mathrm T}[\ell]]
=
L\mathbb{E}[\mathbf s\mathbf s^{\mathrm T}]
=
\mathbf 0$; for sufficiently large $L$, $\mathbf S\mathbf S^{\mathrm T}\approx \mathbf 0$ holds, which leads to
\begin{equation}
\begin{aligned}
\label{uttvtrace}
\tilde u
&=
\mathrm{tr}
\left\{
\mathbf X^{\mathrm T}
\mathbf Q^{\mathrm T}
\mathbf D_{b,r}^{\mathrm T}
\mathbf R_w^{-1}
\mathbf D_{b,r}
\mathbf Q
\mathbf X
\right\}\\
&=
\mathrm{tr}
\left\{
\mathbf Q
\mathbf W \mathbf S 
\mathbf S^{\mathrm T} {\mathbf W}^{\mathrm T}
\mathbf Q^{\mathrm T}
\mathbf D_{b,r}^{\mathrm T}
\mathbf R_w^{-1}
\mathbf D_{b,r}
\right\}=0,\\
\tilde t
&=
\mathrm{tr}
\left\{
\mathbf X^{\mathrm T}
\dot{\mathbf Q}^{\mathrm T}
\mathbf D_{b,r}^{\mathrm T}
\mathbf R_w^{-1}
\mathbf D_{b,r}
\mathbf Q
\mathbf X
\right\}\\
&=
\mathrm{tr}
\left\{
\mathbf Q
\mathbf W \mathbf S 
\mathbf S^{\mathrm T} {\mathbf W}^{\mathrm T}
\dot{\mathbf Q}^{\mathrm T}
\mathbf D_{b,r}^{\mathrm T}
\mathbf R_w^{-1}
\mathbf D_{b,r}
\right\}=0,\\
\tilde v
&=
\mathrm{tr}
\left\{
\mathbf X^{\mathrm T}
\dot{\mathbf Q}^{\mathrm T}
\mathbf D_{b,r}^{\mathrm T}
\mathbf R_w^{-1}
\mathbf D_{b,r}
\dot{\mathbf Q}
\mathbf X
\right\}\\
&=
\mathrm{tr}
\left\{
\dot{\mathbf Q}
\mathbf W \mathbf S 
\mathbf S^{\mathrm T} {\mathbf W}^{\mathrm T}
\dot{\mathbf Q}^{\mathrm T}
\mathbf D_{b,r}^{\mathrm T}
\mathbf R_w^{-1}
\mathbf D_{b,r}
\right\}=0.
\end{aligned}
\end{equation}
Substituting $\tilde u=\tilde t=\tilde v=0$ into~\eqref{abcpqr}, we obtain
\begin{equation}
\begin{aligned}
&A=\frac{u}{2},~
B=\frac{u}{2},~
C=0,\\
&p=\frac{\Re\{\alpha^{*}t\}}{2},~
q=-\frac{\Im\{\alpha^{*}t\}}{2},~
r=\frac{|\alpha|^2 v}{2}.
\label{eq:ABC_simplified}
\end{aligned}
\end{equation}
Accordingly, the denominator term in~\eqref{eq:CRB_theta_rare_closed_re} becomes $AB-C^2
=
\frac{u^2}{4}$, while the numerator is simplified as
\begin{equation}
\begin{aligned}
Bp^2-2Cpq+Aq^2
&=
\frac{u}{2}\left(\frac{\Re\{\alpha^{*}t\}}{2}\right)^2
+
\frac{u}{2}\left(\frac{\Im\{\alpha^{*}t\}}{2}\right)^2 \\
&=
\frac{u}{8}
\left(
\Re^2\{\alpha^{*}t\}+\Im^2\{\alpha^{*}t\}
\right) \\
&=
\frac{u}{8}|\alpha^{*}t|^2
=
\frac{u}{8}|\alpha|^2|t|^2.
\end{aligned}
\label{eq:num_simplified}
\end{equation}
Therefore,
\begin{equation}
\frac{Bp^2-2Cpq+Aq^2}{AB-C^2}
=
\frac{\frac{u}{8}|\alpha|^2|t|^2}{\frac{u^2}{4}}
=
\frac{|\alpha|^2|t|^2}{2u}.
\label{eq:schur_simplified}
\end{equation}
Substituting~\eqref{eq:ABC_simplified} and~\eqref{eq:schur_simplified} into~\eqref{eq:CRB_theta_rare_closed_re} yields
\begin{equation}
\mathrm{CRB}_{\theta}(\mathbf W,\boldsymbol\Phi)
=
\frac{2}{|\alpha|^2}
\left(
v-\frac{|t|^2}{u}
\right)^{-1}.
\label{eq:CRB_theta_final_simplified}
\end{equation}
Finally, by substituting~\eqref{utvtrace} into~\eqref{eq:CRB_theta_final_simplified}, $\mathrm{CRB}_{\theta}(\mathbf W,\boldsymbol\Phi)$ becomes
\begin{equation}
\begin{aligned}
\mathrm{CRB}_{\theta}(\mathbf W,\boldsymbol\Phi)
=&
\frac{2}{L|\alpha|^2}
\Big(
\mathrm{tr}
\Big\{
\dot{\mathbf Q}\mathbf W\mathbf W^{*}\dot{\mathbf Q}^{*}
\mathbf D_{b,r}^{*}\mathbf R_w^{-1}\mathbf D_{b,r}
\Big\}\\
&-
\frac{
\left|
\mathrm{tr}
\left\{
\dot{\mathbf Q}\mathbf W\mathbf W^{*}\mathbf Q^{*}
\mathbf D_{b,r}^{*}\mathbf R_w^{-1}\mathbf D_{b,r}
\right\}
\right|^2
}{
\mathrm{tr}
\left\{
\mathbf Q\mathbf W\mathbf W^{*}\mathbf Q^{*}
\mathbf D_{b,r}^{*}\mathbf R_w^{-1}\mathbf D_{b,r}
\right\}
}
\Big)^{-1}.
\label{eq:CRB_theta_trace_final}
\end{aligned}
\end{equation}

\section{Problem Formulation}
\label{subsec:prob_formulation}
\subsection{SINR of RARE}
Define the LO-phase-aligned effective channel $\mathbf{E}_k(\boldsymbol{\Phi})
\triangleq
\mathbf{D}_{b,k}\mathbf{H}_{\mathrm{eff},k}(\boldsymbol{\Phi})
\in\mathbb{C}^{M\times N_t}$. Then~\eqref{eq:RAR_real_model2} can be rewritten as
\begin{equation}
{\mathbf{z}}_k
=
\Re\left\{
\mathbf{E}_k(\boldsymbol{\Phi})\mathbf{W}\mathbf{s}
\right\}
+\tilde{\mathbf{n}}_k
+\boldsymbol{\eta}_k,
\label{eq:zk_real_compact}
\end{equation}
where $\tilde{\mathbf{n}}_k
\triangleq
\Re\left\{
\mathbf{D}_{b,k}\mathbf{H}_{RU,k}\boldsymbol{\Phi}\mathbf{z}_0
\right\}\in\mathbb{R}^{M}$. Define the effective field vectors $\mathbf{g}_{k,i}(\mathbf{W},\boldsymbol{\Phi})
\triangleq
\mathbf{E}_k(\boldsymbol{\Phi})\mathbf{w}_i
\in\mathbb{C}^{M}~( i=1,\cdots,K)$, and stack them as $\mathbf{G}_k(\mathbf{W},\boldsymbol{\Phi})
\triangleq
\mathbf{E}_k(\boldsymbol{\Phi})\mathbf{W}
=
\big[\mathbf{g}_{k,1} \cdots \mathbf{g}_{k,K}\big]\in\mathbb{C}^{M\times K}$. Hence, the observation becomes
\begin{equation}
{\mathbf{z}}_k
=
\Re\left\{
\mathbf{g}_{k,k}s_k
\right\}
+
\sum_{i\neq k}\Re\left\{
\mathbf{g}_{k,i}s_i
\right\}
+\tilde{\mathbf{n}}_k
+\boldsymbol{\eta}_k.
\label{eq:zk_signal_interf}
\end{equation}
Since $\{s_i\}$ are unit-power proper signal: $\mathbb{E}\left\{\big\|\Re\left\{
\mathbf{g}_{k,k}s_k
\right\}\big\|_2^2\right\}
=
\frac{1}{2}\|\mathbf{g}_{k,k}\|_2^2$. Moreover, since $\mathbf{z}_0\sim\mathcal{CN}(\mathbf{0},\sigma_z^2\mathbf{I}_N)$ is circularly symmetric, $\tilde{\mathbf{n}}_k$ satisfies $\mathbb{E}\left\{\big\|\tilde{\mathbf{n}}_k\big\|_2^2\right\}
=
\frac{\sigma_z^2}{2}\left\|\mathbf{D}_{b,k}\mathbf{H}_{RU,k}\boldsymbol{\Phi}\right\|_F^2$, and the readout noise contributes $\mathbb{E}\left\{\big\|\boldsymbol{\eta}_k\big\|_2^2\right\}
=
\mathrm{tr}\left(\frac{\sigma_{q,k}^2}{2}\mathbf{I}_M\right)
=
\frac{M\sigma_{q,k}^2}{2}$. Motivated by these, we define an SINR that is consistent with the real-part detection model of RARE as
\begin{equation}
\mathrm{SINR}_k(\mathbf{W},\boldsymbol{\Phi})
\triangleq
\frac{
\|\mathbf{g}_{k,k}\|_2^2
}{
\sum_{i\neq k}\|\mathbf{g}_{k,i}\|_2^2
+
\sigma_z^2\left\|\mathbf{D}_{b,k}\mathbf{H}_{RU,k}\boldsymbol{\Phi}\right\|_F^2
+
M\sigma_{q,k}^2
}.
\label{eq:SINR_RAR_def_rig}
\end{equation}
\subsection{ARIS Power Consumption Model}
The ARIS power consumption consists of four parts: i) the amplified incident BS signal power, ii) the target-related re-amplified echo component, iii) the ARIS noise coupled through the target-related loop, and iv) the intrinsic amplifier-noise power, which can be respectively modeled as
\begin{equation}
\begin{aligned}
&P_{\mathrm{RIS}}(\mathbf{W},\boldsymbol{\Phi})\\
&=
\mathbb E\left[\left\|
\boldsymbol{\Phi}(\mathbf{H}_{BR}\mathbf x+\mathbf z_0)
\right\|_2^2\right]\\
&~~~~+\mathbb E\left[\left\|
\boldsymbol{\Phi}\big(\alpha \mathbf h_{r,t}\mathbf h_{r,t}^{\mathrm T} \boldsymbol{\Phi}(\mathbf{H}_{BR}\mathbf x+\mathbf z_0)+\mathbf z_1\big)
\right\|_2^2\right]\\
&=\|\boldsymbol{\Phi}\mathbf{H}_{BR}\mathbf{W}\|_F^2+\sigma_{\alpha}^{2}\left\|\boldsymbol{\Phi}\mathbf{h}_{r,t}(\theta)\mathbf{h}_{r,t}^{\mathrm{T}}(\theta)\boldsymbol{\Phi}\mathbf{H}_{BR}\mathbf{W}\right\|_F^2\\
&~~~~+\sigma_{\alpha}^{2}\sigma_z^{2}\left\|
\boldsymbol{\Phi}\mathbf{h}_{r,t}(\theta)\mathbf{h}_{r,t}^{\mathrm{T}}(\theta)\boldsymbol{\Phi}
\right\|_F^2+2\sigma_z^2\|\boldsymbol{\Phi}\|_F^2 ,
\label{eq:PRIS_model}
\end{aligned}
\end{equation}
where $\sigma_{\alpha}^{2}\triangleq \mathbb{E}[|\alpha|^2]$ denotes the power of the target reflection coefficient. 

In addition, let $P_c$ and $P_{\mathrm{DC}}$ denote the circuit power consumption~\cite{aris5}: i) $P_c$, representing the logical control and switching power required for each element, and ii) $P_{\mathrm{DC}}$, representing the DC bias power required to support active reflection. Then, with $\xi_e=\frac{1}{v}$ with efficiency $v\in(0, 1)$, the total ARIS power consumption is expressed as $N P_c+N P_{\mathrm{DC}}+ \xi_e P_{\mathrm{RIS}}(\mathbf{W},\boldsymbol{\Phi})$, constrained by the budget $P_{\max,a,t}$. Therefore, the constraint with respect to $P_{\mathrm{RIS}}$ is
\begin{equation}
P_{\mathrm{RIS}}(\mathbf{W},\boldsymbol{\Phi})\le v(P_{\max,a,t}-NP_c -NP_{\mathrm{DC}})\triangleq P_{\max,a}.
\label{eq:P_constraint}
\end{equation}

\subsection{CRB-Minimization Problem}
We aim to minimize $\mathrm{CRB}_{\theta}$ in~\eqref{eq:CRB_theta_trace_final} subject to the power budgets, user SINR, and ARIS amplitude constraints:
\begin{equation}
\begin{aligned}
\label{p1}
&\min_{\mathbf{W},\boldsymbol{\Phi}}~\mathrm{CRB}_{\theta}(\mathbf{W},\boldsymbol{\Phi})
\\
\mathrm{s.t.}~&\mathrm{tr}(\mathbf{W}\mathbf{W}^*)\le P_{\max},~ P_{\mathrm{RIS}}(\mathbf{W},\boldsymbol{\Phi})\le P_{\max, a},
\\
& \mathrm{SINR}_k(\mathbf{W},\boldsymbol{\Phi})\ge \gamma_k~(\forall k),~ 0<a_n\le a_{\max}~(\forall n).
\end{aligned}
\end{equation}
Problem~\eqref{p1} is highly non-convex due to the fractional CRB structure, the coupled SINR constraints, and the high-order dependence of the ARIS power on $(\mathbf W,\boldsymbol\Phi)$.  To address this challenge, we develop an AO framework that iteratively optimizes $\mathbf W$ and $\boldsymbol\Phi$. 

\section{Proposed Framework}
\subsection{Design of $\mathbf W$ for fixed $\boldsymbol\Phi$}
For any fixed $\boldsymbol\Phi$, we first reformulate the objective in an equivalent maximization form with respect to $\mathbf W$:
\begin{equation}
\label{eq:fW}
\begin{aligned}
f(\mathbf W)\triangleq&\mathrm{tr}\left(\dot{\mathbf Q}\mathbf W\mathbf W^*\dot{\mathbf Q}^*\mathbf D_{b,r}^*\mathbf R_w^{-1}\mathbf D_{b,r}\right)\\
&-\frac{\left|\mathrm{tr}\left(\dot{\mathbf Q}\mathbf W\mathbf W^*\mathbf Q^*\mathbf D_{b,r}^*\mathbf R_w^{-1}\mathbf D_{b,r}\right)\right|^2}{\mathrm{tr}\left(\mathbf Q\mathbf W\mathbf W^*\mathbf Q^*\mathbf D_{b,r}^*\mathbf R_w^{-1}\mathbf D_{b,r}\right)}.
\end{aligned}
\end{equation}
To facilitate tractable optimization, we introduce the beamforming covariance matrices $\{\mathbf W_k\triangleq\mathbf w_k \mathbf w_k^*\}_{k=1}^K$, and define $\mathbf R_W \triangleq \sum_{k=1}^K \mathbf W_k= \mathbf W\mathbf W^*$. Then for $\mathbf H_k \triangleq \mathbf E_k^* \mathbf E_k$, the SINR constraint of user $k$ in~\eqref{p1} becomes
\begin{equation}
\label{eq:sinr_linear}
\mathrm{tr}(\mathbf H_k \mathbf W_k)\ge\gamma_k\left(\sum_{i\neq k}\mathrm{tr}(\mathbf H_k \mathbf W_i)+\sigma_k^2(\boldsymbol\Phi)
\right),
\end{equation}
where $\sigma_k^2(\boldsymbol\Phi)\triangleq\sigma_{q,k}^2 M+\sigma_z^2
\left\|
\mathbf D_{b,k}\mathbf H_{RU,k}\boldsymbol\Phi
\right\|_F^2$ is the equivalent noise power. The transmit power constraint in~\eqref{p1} becomes $\sum_{k=1}^K \mathrm{tr}(\mathbf W_k) \le P_{\max}$. Next, we introduce an auxiliary variable $t_W$ and rewrite the objective in epigraph form as $f(\mathbf W) \ge t_W$. The ARIS power constraint in~\eqref{p1} can be rewritten as
\begin{equation}
\mathrm{tr}(\mathbf E_\Phi \mathbf R_W) + c_\Phi \le P_{\max,a},
\label{eq:aris_power_w}
\end{equation}
where $\mathbf E_\Phi
\triangleq
\mathbf H_{BR}^*\boldsymbol\Phi^*\boldsymbol\Phi\mathbf H_{BR}
+\sigma_\alpha^2
\mathbf H_{BR}^*
\boldsymbol\Phi^*
(\mathbf h_{r,t}\mathbf h_{r,t}^{\mathrm T})^*
\boldsymbol\Phi^*
\boldsymbol\Phi
\mathbf h_{r,t}\mathbf h_{r,t}^{\mathrm T}
\boldsymbol\Phi
\mathbf H_{BR}$ and $c_\Phi
\triangleq
\sigma_\alpha^2\sigma_z^2
\left\|
\boldsymbol\Phi\mathbf h_{r,t}\mathbf h_{r,t}^{\mathrm T}\boldsymbol\Phi
\right\|_F^2
+
2\sigma_z^2\|\boldsymbol\Phi\|_F^2$. Therefore, the $\mathbf W$-subproblem can be formulated as
\begin{equation}
\begin{aligned}
\label{p1w}
&\max_{\{\mathbf W_k\}, t_W}~t_W\\
\mathrm{s.t.}~
&f(\mathbf W) \ge t_W,~\sum_{k=1}^K \mathrm{tr}(\mathbf W_k) \le P_{\max},\\
&\eqref{eq:sinr_linear},~\eqref{eq:aris_power_w},~\mathbf W_k \succeq \mathbf 0,~\mathrm{rank}(\mathbf W_k)=1~(\forall k).
\end{aligned}
\end{equation}
Problem~\eqref{p1w} is still non-convex due to the rank-one constraints and the fractional structure in $f(\mathbf W)$. In the following, we apply SDR to drop the rank-one constraints and transform the fractional constraint into a linear matrix inequality (LMI) using the Schur complement, resulting in a convex SDP that can be efficiently solved~\cite{sdr, boyd}.

We first reformulate $f(\mathbf W)\ge t_W$ into an equivalent LMI. For convenience, define $A(\mathbf W) \triangleq \mathrm{tr}(\mathbf A_\Phi \mathbf R_W),~B(\mathbf W) \triangleq \mathrm{tr}(\mathbf B_\Phi \mathbf R_W),~C(\mathbf W) \triangleq \mathrm{tr}(\mathbf C_\Phi \mathbf R_W)$, where
\begin{equation}
\begin{aligned}
&\mathbf A_\Phi \triangleq \dot{\mathbf Q}^*(\boldsymbol\Phi,\theta) \mathbf M \dot{\mathbf Q}(\boldsymbol\Phi,\theta),~
\mathbf B_\Phi \triangleq \mathbf Q^*(\boldsymbol\Phi,\theta) \mathbf M \dot{\mathbf Q}(\boldsymbol\Phi,\theta),\\
&\mathbf C_\Phi \triangleq \mathbf Q^*(\boldsymbol\Phi,\theta) \mathbf M \mathbf Q(\boldsymbol\Phi,\theta),
\end{aligned}
\end{equation}
with $\mathbf M \triangleq \mathbf D_{b,r}^*\mathbf R_w^{-1}\mathbf D_{b,r}$. Then, $f(\mathbf W)\ge t_W$ can be equivalently written as
\begin{equation}
\label{eq:quad_ineq}
A(\mathbf W) - \frac{|B(\mathbf W)|^2}{C(\mathbf W)} \ge t_W\leftrightarrow|B(\mathbf W)|^2 \le (A(\mathbf W)-t_W) C(\mathbf W),
\end{equation}
provided that $C(\mathbf W) > 0$. By applying the Schur complement,~\eqref{eq:quad_ineq} is equivalently expressed as the following LMI: $\begin{bmatrix}
C(\mathbf W) & B(\mathbf W)\\
B(\mathbf W)^* & A(\mathbf W)-t_W
\end{bmatrix}
\succeq \mathbf 0$. Consequently, by dropping the non-convex rank-one constraints $\mathrm{rank}(\mathbf W_k)=1$, the $\mathbf W$-subproblem can be relaxed into:
\begin{equation}
\begin{aligned}
&\max_{\{\mathbf W_k\}, t_W}~t_W\\
\mathrm{s.t.}~
&\begin{bmatrix}
C(\mathbf W) & B(\mathbf W)\\
B(\mathbf W)^* & A(\mathbf W)-t_W
\end{bmatrix}
\succeq \mathbf 0,~\sum_{k=1}^K \mathrm{tr}(\mathbf W_k) \le P_{\max},\\
&\mathrm{tr}(\mathbf E_\Phi \mathbf R_W) + c_\Phi \le P_{\max,a},~\mathbf W_k \succeq \mathbf 0~(\forall k),\\
&\mathrm{tr}(\mathbf H_k \mathbf W_k)
\ge\gamma_k\left(\sum_{i\neq k}\mathrm{tr}(\mathbf H_k \mathbf W_i)+\sigma_k^2(\boldsymbol\Phi)
\right)~(\forall k),
\label{prbsdp}
\end{aligned}
\end{equation}
which is a convex SDP and can be efficiently solved using CVX~\cite{boyd}. Let $\{\mathbf W_k^\star\}$ denote the optimal solution. For $k$ with $\mathrm{rank}(\mathbf W_k^\star)=1$, the corresponding beamforming vectors can be directly obtained via eigenvalue decomposition as $\mathbf w_k^\star = \sqrt{\lambda_{\max}(\mathbf W_k^\star)}\mathbf u_{\max}(\mathbf W_k^\star)$~\cite{crbaris}, where $\lambda_{\max}(\mathbf W_k^\star)$ denotes the maximum eigenvalue of $\mathbf W_k^\star$, and $\mathbf u_{\max}(\mathbf W_k^\star)$ is the corresponding unit-norm eigenvector. Otherwise, a feasible rank-one solution can be constructed from corresponding $\{\mathbf W_k^\star\}$ using Gaussian randomization~\cite{sdr}.

\subsection{Design of $\boldsymbol\Phi$ for Fixed $\mathbf W$}
\label{subsec:phi_design}
For fixed $\mathbf W$, the optimization variable is the ARIS coefficient vector $\boldsymbol\phi \triangleq [\phi_1 \cdots \phi_N]^{\mathrm T}\in\mathbb C^N,~\boldsymbol\Phi=\mathrm{diag}(\boldsymbol\phi)$. Recalling~\eqref{eq:CRB_theta_final_simplified}-\eqref{eq:CRB_theta_trace_final}, minimizing $\mathrm{CRB}_{\theta}(\mathbf W,\boldsymbol\Phi)$ is equivalent to maximizing $g(\boldsymbol\phi)\triangleq v(\boldsymbol\phi)-\frac{|t(\boldsymbol\phi)|^2}{u(\boldsymbol\phi)}$ subject to the constraints.
\subsubsection{Lifted Representation of the Objective}
Let $\mathbf h \triangleq \mathbf h_{r,t}(\theta),~
\dot{\mathbf h} \triangleq \frac{\partial \mathbf h_{r,t}(\theta)}{\partial\theta}$, and define $\mathbf B_0 \triangleq \mathrm{diag}(\mathbf h)\mathbf H_{BR}\in\mathbb C^{N\times N_t},~\tilde{\mathbf B}_0\triangleq\mathrm{diag}(\mathbf h)\mathbf H_{RB,r}\in\mathbb C^{N\times N_t}~\mathbf B_1 \triangleq \mathrm{diag}(\dot{\mathbf h})\mathbf H_{BR}\in\mathbb C^{N\times N_t},~\tilde{\mathbf B}_1\triangleq\mathrm{diag}(\dot{\mathbf h})\mathbf H_{RB,r}\in\mathbb C^{N\times N_t}$. Then, $\mathbf Q(\boldsymbol\Phi,\theta)$ can be written as
\begin{equation}
\mathbf Q(\boldsymbol\Phi,\theta)
=
\tilde{\mathbf B}_0^{\mathrm T}\boldsymbol\phi\boldsymbol\phi^{\mathrm T}\mathbf B_0,
\label{eq:Q_phi_exact}
\end{equation}
and its derivative with respect to $\theta$ is
\begin{equation}
\dot{\mathbf Q}(\boldsymbol\Phi,\theta)=\tilde{\mathbf B}_1^{\mathrm T}\boldsymbol\phi\boldsymbol\phi^{\mathrm T}\mathbf B_0+\tilde{\mathbf B}_0^{\mathrm T}\boldsymbol\phi\boldsymbol\phi^{\mathrm T}\mathbf B_1.
\label{eq:Qdot_phi_exact}
\end{equation}
Now define the lifted variable $\mathbf x \triangleq \mathrm{vec}(\boldsymbol\phi\boldsymbol\phi^{\mathrm T})=\boldsymbol\phi\otimes\boldsymbol\phi\in\mathbb C^{N^2}$, and let $\mathbf M_r \triangleq \mathbf D_{b,r}^*\mathbf R_w^{-1}\mathbf D_{b,r}$. 
Then there exist $\mathbf T\in\mathbb C^{N^2\times N^2}$ and positive definite matrices $\mathbf U,\mathbf V\in\mathbb C^{N^2\times N^2}$ such that
\begin{equation}
u(\boldsymbol\phi)=\mathbf x^{*}\mathbf U\mathbf x,~
t(\boldsymbol\phi)=\mathbf x^{*}\mathbf T\mathbf x,~
v(\boldsymbol\phi)=\mathbf x^{*}\mathbf V\mathbf x,
\label{eq:utv_x_form}
\end{equation}
where $\mathbf U
\triangleq
L\mathbf G_0^{*}
(\mathbf M_r^{\mathrm T}\otimes \mathbf R_W)
\mathbf G_0,~
\mathbf T
\triangleq
L\mathbf G_1^{*}
(\mathbf M_r^{\mathrm T}\otimes \mathbf R_W)
\mathbf G_0,~\mathbf V
\triangleq
L\mathbf G_1^{*}
(\mathbf M_r^{\mathrm T}\otimes \mathbf R_W)
\mathbf G_1$ with $\mathbf G_0 \triangleq {\mathbf B}_0^{\mathrm T}\otimes \tilde{\mathbf B}_0^{\mathrm T},~\mathbf G_1 \triangleq {\mathbf B}_0^{\mathrm T}\otimes \tilde{\mathbf B}_1^{\mathrm T}+{\mathbf B}_1^{\mathrm T}\otimes \tilde{\mathbf B}_0^{\mathrm T}$. The derivation is given in Appendix~\ref{app:lifting_utv}. Hence,~$g(\boldsymbol\phi)$ can be equivalently written as
\begin{equation}
g(\boldsymbol\phi)=\mathbf x^{*}\mathbf V\mathbf x-\frac{\left|\mathbf x^{*}\mathbf T\mathbf x\right|^2}
{\mathbf x^{*}\mathbf U\mathbf x}.
\label{eq:g_phi_x_form}
\end{equation}
\subsubsection{Quadratic Transform of the Objective}
Since $u>0$, the fractional term in~\eqref{eq:g_phi_x_form} can be handled via the quadratic transform~\cite{fracp}: $-\frac{|t|^2}{u}=\min_{y\in\mathbb C}\left(u|y|^2-2\Re\{y^*t\}\right)$, whose minimizer is $y^\star=\frac{t}{u}$. Therefore, maximizing~\eqref{eq:g_phi_x_form} is equivalent to maximizing
\begin{equation}
\min_{y\in\mathbb C}~
\mathbf x^{*}\mathbf V\mathbf x+|y|^2\mathbf x^{*}\mathbf U\mathbf x-2\Re\left\{y^* \mathbf x^{*}\mathbf T\mathbf x
\right\}
\label{eq:maxmin_phi_y}
\end{equation}
with respect to $\boldsymbol\phi$ satisfying the constraints. At the $s$th inner iteration, for fixed $\boldsymbol\phi^{(s)}$, the auxiliary scalar $y$ is updated as: $y^{(s)}=\frac{\mathbf x^{(s)*}\mathbf T\mathbf x^{(s)}}{\mathbf x^{(s)*}\mathbf U\mathbf x^{(s)}},~\mathbf x^{(s)}=\boldsymbol\phi^{(s)}\otimes\boldsymbol\phi^{(s)}$. Then, with $y^{(s)}$ fixed, the $\boldsymbol\phi$-update reduces to maximizing $f_y(\boldsymbol\phi)\triangleq\mathbf x^{*}\mathbf C_y^{(s)}\mathbf x$ by~\eqref{eq:maxmin_phi_y}, where $\mathbf C_y^{(s)}\triangleq \mathbf V+|y^{(s)}|^2\mathbf U-y^{(s)*}\mathbf T-y^{(s)}\mathbf T^{*}$.
\subsubsection{MM Surrogate of the Objective}
At the $s$th MM iteration, let $\boldsymbol\phi^{(s)}$ be the current feasible point and define $\mathbf x^{(s)}=\boldsymbol\phi^{(s)}\otimes\boldsymbol\phi^{(s)}$. For fixed $y^{(s)}$, maximizing $f_y(\boldsymbol\phi)$ is equivalent to minimizing $-\mathbf x^{*}\mathbf C_y^{(s)}\mathbf x$. Define the Hermitian part $\mathbf H_y^{(s)} \triangleq \frac{\mathbf C_y^{(s)}+\mathbf C_y^{(s)*}}{2}$, so that $\mathbf x^{*}\mathbf C_y^{(s)}\mathbf x
=
\mathbf x^{*}\mathbf H_y^{(s)}\mathbf x$. Choose any constant $\lambda_y^{(s)}\ge \lambda_{\max}(\mathbf H_y^{(s)})$. Then, by the standard quadratic majorization~\cite{mmtsp, mmtit},
\begin{equation}
\begin{aligned}
&-\mathbf x^{*}\mathbf H_y^{(s)}\mathbf x
\le
-\mathbf x^{(s)*}\mathbf H_y^{(s)}\mathbf x^{(s)}-2\Re\left\{
\mathbf x^{(s)*}\mathbf H_y^{(s)}(\mathbf x-\mathbf x^{(s)})
\right\}\\
&~~~~~~~~~~~~~~~~~~+\lambda_y^{(s)}\|\mathbf x-\mathbf x^{(s)}\|_2^2\\
&\rightarrow-\mathbf x^{*}\mathbf H_y^{(s)}\mathbf x
\le
\lambda_y^{(s)}\|\mathbf x\|_2^2
-
2\Re\left\{
\mathbf x^{(s)*}\big(\mathbf H_y^{(s)}+\lambda_y^{(s)}\mathbf I\big)\mathbf x
\right\}
+
c_y^{(s)},
\label{eq:mm_obj_x_expand}
\end{aligned}
\end{equation}
where $c_y^{(s)}
\triangleq
\mathbf x^{(s)*}\mathbf H_y^{(s)}\mathbf x^{(s)}
+\lambda_y^{(s)}\|\mathbf x^{(s)}\|_2^2$ is independent of $\boldsymbol\phi$. Now define $\mathbf q_y^{(s)}
\triangleq
\big(\mathbf H_y^{(s)}+\lambda_y^{(s)}\mathbf I\big)\mathbf x^{(s)},
~
\mathbf Q_y^{(s)}
\triangleq
\mathrm{unvec}_{N\times N}\big(\mathbf q_y^{(s)}\big)$, where $\mathrm{unvec}_{N\times N}(\cdot)$ reshapes a vector into an $N\times N$ matrix. Then $\mathbf q_y^{(s)*}\mathbf x
=
\mathrm{vec}(\mathbf Q_y^{(s)})^{*}\mathrm{vec}(\boldsymbol\phi\boldsymbol\phi^{\mathrm T})
=
\boldsymbol\phi^{\mathrm T}\mathbf Q_y^{(s)*}\boldsymbol\phi$, and moreover, $\|\mathbf x\|_2^2
=
\|\boldsymbol\phi\otimes\boldsymbol\phi\|_2^2
=
(\boldsymbol\phi^*\boldsymbol\phi)^2
\le
N a_{\max}^2\boldsymbol\phi^*\boldsymbol\phi$. Substituting them into~\eqref{eq:mm_obj_x_expand}, we obtain
\begin{equation}
-\mathbf x^{*}\mathbf C_y^{(s)}\mathbf x
\le
\lambda_y^{(s)} N a_{\max}^2\boldsymbol\phi^*\boldsymbol\phi
-
2\Re\left\{
\boldsymbol\phi^{\mathrm T}\mathbf Q_y^{(s)*}\boldsymbol\phi
\right\}
+
c_y^{(s)}.
\label{eq:mm_obj_phi_quad_fixed}
\end{equation}
The remaining term $\Re\{\boldsymbol\phi^{\mathrm T}\mathbf Q_y^{(s)*}\boldsymbol\phi\}$ in~\eqref{eq:mm_obj_phi_quad_fixed} is still non-convex in general. To handle it, introduce the real-valued embedding $\bar{\boldsymbol\phi}
\triangleq
\begin{bmatrix}
\Re\{\boldsymbol\phi\}\\
\Im\{\boldsymbol\phi\}
\end{bmatrix},
~
\bar{\mathbf Q}_y^{(s)}
\triangleq
\begin{bmatrix}
\Re\{\mathbf Q_y^{(s)}\} & -\Im\{\mathbf Q_y^{(s)}\}\\
\Im\{\mathbf Q_y^{(s)}\} & \Re\{\mathbf Q_y^{(s)}\}
\end{bmatrix}$. Then $\Re\left\{
\boldsymbol\phi^{\mathrm T}\mathbf Q_y^{(s)*}\boldsymbol\phi
\right\}
=
\bar{\boldsymbol\phi}^{\mathrm T}\bar{\mathbf Q}_y^{(s)}\bar{\boldsymbol\phi}$. Choose any $\mu_y^{(s)}\ge \lambda_{\max}\big(\bar{\mathbf Q}_y^{(s)}+\bar{\mathbf Q}_y^{(s)\mathrm T}\big)$. Then, by the second-order upper-bound:
\begin{equation}
\begin{aligned}
\bar{\boldsymbol\phi}^{\mathrm T}\bar{\mathbf Q}_y^{(s)}\bar{\boldsymbol\phi}
\le&
\bar{\boldsymbol\phi}^{(s)\mathrm T}\bar{\mathbf Q}_y^{(s)}\bar{\boldsymbol\phi}^{(s)}+
\bar{\boldsymbol\phi}^{(s)\mathrm T}
\big(
\bar{\mathbf Q}_y^{(s)}+\bar{\mathbf Q}_y^{(s)\mathrm T}
\big)
(\bar{\boldsymbol\phi}-\bar{\boldsymbol\phi}^{(s)})\\
&+
\frac{\mu_y^{(s)}}{2}
\|\bar{\boldsymbol\phi}-\bar{\boldsymbol\phi}^{(s)}\|_2^2.
\end{aligned}
\label{eq:second_taylor_obj_fixed}
\end{equation}
Substituting~\eqref{eq:second_taylor_obj_fixed} into~\eqref{eq:mm_obj_phi_quad_fixed}, we obtain
\begin{equation}
-\mathbf x^{*}\mathbf C_y^{(s)}\mathbf x
\le
\boldsymbol\phi^*\mathbf A_y^{(s)}\boldsymbol\phi
+
\Re\left\{
\mathbf b_y^{(s)*}\boldsymbol\phi
\right\}
+
c_{y,0}^{(s)},
\label{eq:objective_surrogate_final_fixed}
\end{equation}
where $\mathbf A_y^{(s)}
\triangleq
\left(\lambda_y^{(s)} N a_{\max}^2+\mu_y^{(s)}\right)\mathbf I_N,~\mathbf b_y^{(s)}
\triangleq
-2\left(
\boldsymbol\ell_{y,\mathrm R}^{(s)}
+
j\boldsymbol\ell_{y,\mathrm I}^{(s)}
\right),~c_{y,0}^{(s)}
\triangleq
c_y^{(s)}
+
2\bar{\boldsymbol\phi}^{(s)\mathrm T}\bar{\mathbf Q}_y^{(s)}\bar{\boldsymbol\phi}^{(s)}
-
\mu_y^{(s)}\|\bar{\boldsymbol\phi}^{(s)}\|_2^2$, with $\boldsymbol\ell_y^{(s)}
\triangleq
\left(
\bar{\mathbf Q}_y^{(s)}+\bar{\mathbf Q}_y^{(s)\mathrm T}-\mu_y^{(s)}\mathbf I_{2N}
\right)\bar{\boldsymbol\phi}^{(s)}
=
\begin{bmatrix}
\boldsymbol\ell_{y,\mathrm R}^{(s)}\\
\boldsymbol\ell_{y,\mathrm I}^{(s)}
\end{bmatrix}$. Herein, $\boldsymbol\ell_{y,\mathrm R}^{(s)}$ and $\boldsymbol\ell_{y,\mathrm I}^{(s)}$ denote the first and last $N$ entries of $\boldsymbol\ell_y^{(s)}\in\mathbb R^{2N}$, respectively.

\subsubsection{Reformulation of the Constraints}
We next reformulate the constraints in terms of $\boldsymbol\phi$.
\paragraph*{\textbf{1) SINR Constraints}}
For fixed $\mathbf W$, define $\mathbf c_{k,i}
\triangleq
\mathbf D_{b,k}\mathbf H_{BU,k}\mathbf w_i\in\mathbb C^M,~\mathbf F_{k,i}
\triangleq
\mathbf D_{b,k}\mathbf H_{RU,k}\mathrm{diag}(\mathbf H_{BR}\mathbf w_i)\in\mathbb C^{M\times N}$. Then $\mathbf g_{k,i}(\mathbf W,\boldsymbol\Phi)
=
\mathbf E_k(\boldsymbol\Phi)\mathbf w_i
=
\mathbf c_{k,i}+\mathbf F_{k,i}\boldsymbol\phi$, which leads to $\|\mathbf g_{k,i}\|_2^2
=
\boldsymbol\phi^*\mathbf A_{k,i}\boldsymbol\phi
+
2\Re\{\mathbf b_{k,i}^*\boldsymbol\phi\}
+
c_{k,i}$, where $\mathbf A_{k,i}\triangleq \mathbf F_{k,i}^*\mathbf F_{k,i},
~
\mathbf b_{k,i}\triangleq \mathbf F_{k,i}^*\mathbf c_{k,i},
~
c_{k,i}\triangleq \|\mathbf c_{k,i}\|_2^2$. Similarly, $\left\|\mathbf D_{b,k}\mathbf H_{RU,k}\boldsymbol\Phi\right\|_F^2
=
\boldsymbol\phi^*\mathbf Z_k\boldsymbol\phi,
~
\mathbf Z_k\triangleq
\mathrm{diag}\left(
\mathrm{diag}\big(
\mathbf H_{RU,k}^*\mathbf H_{RU,k}
\big)
\right)$. Hence, the SINR constraint is equivalent to
\begin{equation}
\begin{aligned}
&
\boldsymbol\phi^*
\Big(
\sum_{i\neq k}\mathbf A_{k,i}
+\sigma_z^2\mathbf Z_k
-\gamma_k^{-1}\mathbf A_{k,k}
\Big)\boldsymbol\phi
\\
&~
+
2\Re\left\{
\Big(
\sum_{i\neq k}\mathbf b_{k,i}
-\gamma_k^{-1}\mathbf b_{k,k}
\Big)^*\boldsymbol\phi
\right\}
+
\sum_{i\neq k}c_{k,i}\\
&-\gamma_k^{-1}c_{k,k}
+
M\sigma_{q,k}^2
\le 0.
\end{aligned}
\label{eq:sinr_quad_final}
\end{equation}
For compactness, define $\mathbf C_k
\triangleq
\sum_{i\neq k}\mathbf A_{k,i}
+\sigma_z^2\mathbf Z_k
-\gamma_k^{-1}\mathbf A_{k,k},~\mathbf d_k
\triangleq
2\left(
\sum_{i\neq k}\mathbf b_{k,i}
-\gamma_k^{-1}\mathbf b_{k,k}
\right),~e_k
\triangleq
\sum_{i\neq k}c_{k,i}
-\gamma_k^{-1}c_{k,k}
+
M\sigma_{q,k}^2$. Then~\eqref{eq:sinr_quad_final} becomes
\begin{equation}
\boldsymbol\phi^*\mathbf C_k\boldsymbol\phi
+
\Re\{\mathbf d_k^*\boldsymbol\phi\}
+
e_k
\le 0~ (\forall k).
\label{eq:sinr_compact_phi}
\end{equation}
However, $\mathbf C_k$ is not guaranteed to be positive semidefinite because of $-\gamma_k^{-1}\mathbf A_{k,k}$. Hence,~\eqref{eq:sinr_compact_phi} is generally non-convex. To expose its structure, we decompose $\mathbf C_k$ as $\mathbf C_k
=
\mathbf C_k^{(+)}
-
\mathbf C_k^{(-)}$, where $\mathbf C_k^{(+)}
\triangleq
\sum_{i\neq k}\mathbf A_{k,i}
+\sigma_z^2\mathbf Z_k
\succeq \mathbf 0$ and $\mathbf C_k^{(-)}
\triangleq
\gamma_k^{-1}\mathbf A_{k,k}
\succeq \mathbf 0$. Substituting it into~\eqref{eq:sinr_compact_phi} yields
\begin{equation}
\boldsymbol\phi^*\mathbf C_k^{(+)}\boldsymbol\phi
-
\boldsymbol\phi^*\mathbf C_k^{(-)}\boldsymbol\phi
+
\Re\{\mathbf d_k^*\boldsymbol\phi\}
+
e_k
\le 0.
\label{eq:sinr_dc_form}
\end{equation}
Now, since $\mathbf C_k^{(-)}\succeq \mathbf 0$, the quadratic function $f_k(\boldsymbol\phi)
\triangleq
\boldsymbol\phi^*\mathbf C_k^{(-)}\boldsymbol\phi$ is convex in $\boldsymbol\phi$. Therefore, its negative $-f_k(\boldsymbol\phi)$ is concave, and can be globally upper-bounded by its first-order expansion at $\boldsymbol\phi^{(s)}$:
\begin{equation}
\begin{aligned}
-\boldsymbol\phi^*\mathbf C_k^{(-)}\boldsymbol\phi
&\le
-\boldsymbol\phi^{(s)*}\mathbf C_k^{(-)}\boldsymbol\phi^{(s)}
-
2\Re\left\{
\boldsymbol\phi^{(s)*}\mathbf C_k^{(-)}(\boldsymbol\phi-\boldsymbol\phi^{(s)})
\right\}\\
&=-2\Re\left\{
\boldsymbol\phi^{(s)*}\mathbf C_k^{(-)}\boldsymbol\phi
\right\}
+
\boldsymbol\phi^{(s)*}\mathbf C_k^{(-)}\boldsymbol\phi^{(s)}.
\end{aligned}
\label{eq:concave_upperbound_step2}
\end{equation}
Substituting~\eqref{eq:concave_upperbound_step2} into~\eqref{eq:sinr_dc_form}, we obtain
\begin{equation}
\begin{aligned}
&
\boldsymbol\phi^*\mathbf C_k^{(+)}\boldsymbol\phi
+
\Re\{\mathbf d_k^*\boldsymbol\phi\}
\\
&~
-2\Re\left\{
\boldsymbol\phi^{(s)*}\mathbf C_k^{(-)}\boldsymbol\phi
\right\}
+
\boldsymbol\phi^{(s)*}\mathbf C_k^{(-)}\boldsymbol\phi^{(s)}
+
e_k
\le 0.
\end{aligned}
\label{eq:sinr_surrogate_expand}
\end{equation} 
Thus, defining $\tilde{\mathbf d}_k^{(s)}
\triangleq
\mathbf d_k
-
2\mathbf C_k^{(-)}\boldsymbol\phi^{(s)}$ and $\tilde e_k^{(s)}
\triangleq
e_k
+
\boldsymbol\phi^{(s)*}\mathbf C_k^{(-)}\boldsymbol\phi^{(s)}$, the convex surrogate of the $k$th SINR constraint, modified from~\eqref{eq:sinr_compact_phi}, is finally written as
\begin{equation}
\boldsymbol\phi^*\mathbf C_k^{(+)}\boldsymbol\phi
+
\Re\{\tilde{\mathbf d}_k^{(s)*}\boldsymbol\phi\}
+
\tilde e_k^{(s)}
\le 0~(\forall k).
\label{eq:sinr_surrogate_final}
\end{equation}

\paragraph*{\textbf{2) ARIS Power Constraint}}
Let $\mathbf H_w \triangleq \mathbf H_{BR}\mathbf W\in\mathbb C^{N\times K},
~
\mathbf p \triangleq \boldsymbol\phi\odot \mathbf h\in\mathbb C^N$. Then the first term of~\eqref{eq:PRIS_model} becomes
\begin{equation}
\|\boldsymbol\Phi\mathbf H_{BR}\mathbf W\|_F^2
=
\boldsymbol\phi^*\mathbf K_1\boldsymbol\phi,
~
\mathbf K_1\triangleq
\mathrm{diag}\big(
\mathrm{diag}(\mathbf H_w\mathbf H_w^*)
\big).
\label{eq:power_term1}
\end{equation}
Next, since $\boldsymbol\Phi\mathbf h\mathbf h^{\mathrm T}\boldsymbol\Phi\mathbf H_{BR}\mathbf W
=
\mathbf p\mathbf p^{\mathrm T}\mathbf H_w$, we have the second term: $\left\|
\boldsymbol\Phi\mathbf h\mathbf h^{\mathrm T}\boldsymbol\Phi\mathbf H_{BR}\mathbf W
\right\|_F^2
=
\|\mathbf p\|_2^2\mathbf p^*\mathbf H_w\mathbf H_w^*\mathbf p$. Herein, since $\mathbf p=\mathrm{diag}(\mathbf h)\boldsymbol\phi$, define $\mathbf J
\triangleq
\mathrm{diag}(\mathbf h)^*
\mathbf H_w\mathbf H_w^*
\mathrm{diag}(\mathbf h),
~
\mathbf D_h
\triangleq
\mathrm{diag}(|h_1|^2,\cdots,|h_N|^2)$, then
\begin{equation}
\left\|
\boldsymbol\Phi\mathbf h\mathbf h^{\mathrm T}\boldsymbol\Phi\mathbf H_{BR}\mathbf W
\right\|_F^2
=
(\boldsymbol\phi^*\mathbf D_h\boldsymbol\phi)
(\boldsymbol\phi^*\mathbf J\boldsymbol\phi).
\label{eq:power_term2}
\end{equation}
Likewise the third term becomes $\left\|
\boldsymbol\Phi\mathbf h\mathbf h^{\mathrm T}\boldsymbol\Phi
\right\|_F^2
=
(\boldsymbol\phi^*\mathbf D_h\boldsymbol\phi)^2$, and for the fourth term: $\|\boldsymbol\Phi\|_F^2=\boldsymbol\phi^*\boldsymbol\phi$. Substituting all into~\eqref{eq:PRIS_model}, the power constraint becomes
\begin{equation}
\begin{aligned}
&\boldsymbol\phi^*\mathbf K_1\boldsymbol\phi
+
\sigma_\alpha^2
(\boldsymbol\phi^*\mathbf D_h\boldsymbol\phi)
(\boldsymbol\phi^*\mathbf J\boldsymbol\phi)\\
&+
\sigma_\alpha^2\sigma_z^2
(\boldsymbol\phi^*\mathbf D_h\boldsymbol\phi)^2
+
2\sigma_z^2\boldsymbol\phi^*\boldsymbol\phi
\le
P_{\max,a}.
\end{aligned}
\label{eq:power_exact_phi}
\end{equation}
We consider the quartic part of~\eqref{eq:power_exact_phi}, which breaks the convexity, given by $p_4(\boldsymbol\phi)
\triangleq
\sigma_\alpha^2
(\boldsymbol\phi^*\mathbf D_h\boldsymbol\phi)
(\boldsymbol\phi^*\mathbf J\boldsymbol\phi)+
\sigma_\alpha^2\sigma_z^2
(\boldsymbol\phi^*\mathbf D_h\boldsymbol\phi)^2$. Define the lifted matrix and vector $\mathbf X_p \triangleq \boldsymbol\phi\boldsymbol\phi^*,~
\mathbf y \triangleq \mathrm{vec}(\mathbf X_p)
= \overline{\boldsymbol\phi}\otimes \boldsymbol\phi$, and let $\mathbf d \triangleq \mathrm{vec}(\mathbf D_h),
~
\mathbf j \triangleq \mathrm{vec}(\mathbf J)$. Then, using $\mathrm{tr}(\mathbf A\mathbf X_p)=\mathrm{vec}(\mathbf A)^*\mathrm{vec}(\mathbf X_p)$, we obtain $\boldsymbol\phi^*\mathbf D_h\boldsymbol\phi
= \mathrm{tr}(\mathbf D_h\mathbf X_p)
= \mathbf d^*\mathbf y$ and $\boldsymbol\phi^*\mathbf J\boldsymbol\phi
= \mathbf j^*\mathbf y$. Since $\mathbf D_h$ and $\mathbf J$ are Hermitian and $\mathbf X_p=\boldsymbol\phi\boldsymbol\phi^*$ is Hermitian and positive semidefinite, $\mathbf d^*\mathbf y, \mathbf j^*\mathbf y\in\mathbb R$ holds. Therefore,
\begin{equation}
(\mathbf d^*\mathbf y)(\mathbf j^*\mathbf y)
=
\mathbf y^*
\left(
\frac{\mathbf d\mathbf j^*+\mathbf j\mathbf d^*}{2}
\right)\mathbf y,
~
(\mathbf d^*\mathbf y)^2
=
\mathbf y^*(\mathbf d\mathbf d^*)\mathbf y.
\label{eq:quartic_to_quad}
\end{equation}
Substituting~\eqref{eq:quartic_to_quad} into $p_4(\boldsymbol\phi)$, we obtain the quadratic form: $p_4(\boldsymbol\phi)
=
\mathbf y^*\mathbf M_p\mathbf y$, where $\mathbf M_p
\triangleq
\sigma_\alpha^2
\frac{\mathbf d\mathbf j^*+\mathbf j\mathbf d^*}{2}
+
\sigma_\alpha^2\sigma_z^2
\mathbf d\mathbf d^*$, and it is readily verified that $\mathbf M_p$ is Hermitian.

Now define for $\boldsymbol\phi^{(s)}$: $\mathbf y^{(s)} = \mathrm{vec}(\boldsymbol\phi^{(s)}\boldsymbol\phi^{(s)*})$. Choose any positive scalar $\lambda_p^{(s)}$ satisfying $\lambda_p^{(s)} \ge \lambda_{\max}(\mathbf M_p)$. Then the following holds with standard quadratic majorization: 
\begin{equation}
\mathbf y^*\mathbf M_p\mathbf y
\le
\lambda_p^{(s)}\|\mathbf y\|_2^2+
2\Re\left\{
\mathbf y^*(\mathbf M_p-\lambda_p^{(s)}\mathbf I)\mathbf y^{(s)}
\right\}
+
c_p^{(s)},
\label{eq:MM_y}
\end{equation}
where $c_p^{(s)}
\triangleq
\mathbf y^{(s)*}
(\lambda_p^{(s)}\mathbf I-\mathbf M_p)
\mathbf y^{(s)}$ is constant with respect to $\boldsymbol\phi$.

Now define $\mathbf q_p^{(s)}
\triangleq
(\mathbf M_p-\lambda_p^{(s)}\mathbf I)\mathbf y^{(s)}$ and $\mathbf Q_p^{(s)}
\triangleq
\mathrm{unvec}_{N\times N}(\mathbf q_p^{(s)})$. Using the identity $\mathrm{vec}(\boldsymbol\phi\boldsymbol\phi^*)^*
\mathrm{vec}(\mathbf Q)
=
\boldsymbol\phi^*\mathbf Q\boldsymbol\phi$, we obtain $\mathbf y^*\mathbf q_p^{(s)}
=
\boldsymbol\phi^*\mathbf Q_p^{(s)}\boldsymbol\phi$ and $\|\mathbf y\|_2^2
=
\|\boldsymbol\phi\boldsymbol\phi^*\|_F^2
=
(\boldsymbol\phi^*\boldsymbol\phi)^2$. Hence,~\eqref{eq:MM_y} becomes
\begin{equation}
p_4(\boldsymbol\phi)
\le
\lambda_p^{(s)}(\boldsymbol\phi^*\boldsymbol\phi)^2
+
2\Re\{\boldsymbol\phi^*\mathbf Q_p^{(s)}\boldsymbol\phi\}
+
c_p^{(s)}.
\label{eq:p4_MM_final}
\end{equation}
However, the quadratic term $2\Re\{\boldsymbol\phi^*\mathbf Q_p^{(s)}\boldsymbol\phi\}$ is not necessarily convex, since $\mathbf Q_p^{(s)}+\mathbf Q_p^{(s)*}$ is not guaranteed to be positive semidefinite. To obtain a convex upper-bound, define $\mathbf H_p^{(s)}
\triangleq
\mathbf Q_p^{(s)}+\mathbf Q_p^{(s)*}$, which is Hermitian, and choose any positive scalar $\mu_p^{(s)}$ satisfying $\mu_p^{(s)}
\ge
\lambda_{\max}(\mathbf H_p^{(s)})$. Then, again by quadratic majorization,
\begin{equation}
\begin{aligned}
\boldsymbol\phi^*\mathbf H_p^{(s)}\boldsymbol\phi
\le
&
\mu_p^{(s)}\boldsymbol\phi^*\boldsymbol\phi
+
2\Re\left\{
\boldsymbol\phi^*
(\mathbf H_p^{(s)}-\mu_p^{(s)}\mathbf I)
\boldsymbol\phi^{(s)}
\right\}
+
d_p^{(s)},
\end{aligned}
\label{eq:Hp_MM}
\end{equation}
where $d_p^{(s)}
\triangleq
\boldsymbol\phi^{(s)*}
(\mu_p^{(s)}\mathbf I-\mathbf H_p^{(s)})
\boldsymbol\phi^{(s)}$ is constant with respect to $\boldsymbol\phi$. Since $2\Re\{\boldsymbol\phi^*\mathbf Q_p^{(s)}\boldsymbol\phi\}
=
\boldsymbol\phi^*\mathbf H_p^{(s)}\boldsymbol\phi$, substituting~\eqref{eq:Hp_MM} into~\eqref{eq:p4_MM_final} yields
\begin{equation}
p_{4}(\boldsymbol\phi)
\le
\lambda_p^{(s)}(\boldsymbol\phi^*\boldsymbol\phi)^2
+
\mu_p^{(s)}\boldsymbol\phi^*\boldsymbol\phi
+
2\Re\left\{
\mathbf b_p^{(s)*}\boldsymbol\phi
\right\}
+
\tilde c_p^{(s)},
\label{eq:p4_MM_final2}
\end{equation}
where $\mathbf b_p^{(s)}
\triangleq
(\mathbf H_p^{(s)}-\mu_p^{(s)}\mathbf I)\boldsymbol\phi^{(s)}$ and $\tilde c_p^{(s)}
\triangleq
c_p^{(s)}+d_p^{(s)}$. Therefore, by substituting~\eqref{eq:p4_MM_final2} into~\eqref{eq:power_exact_phi}, we obtain the convex MM surrogate of the ARIS power constraint:
\begin{equation}
\begin{aligned}
&\lambda_p^{(s)}(\boldsymbol\phi^*\boldsymbol\phi)^2
+
\boldsymbol\phi^*(\mathbf K_1+2\sigma_z^2\mathbf I+\mu_p^{(s)}\mathbf I)\boldsymbol\phi
\\
&+2\Re\left\{
\mathbf b_p^{(s)*}\boldsymbol\phi
\right\}
+
\tilde c_p^{(s)}
\le
P_{\max,a}.
\end{aligned}
\label{eq:power_MM_final_convex}
\end{equation}

\subsubsection{Convex MM $\boldsymbol\Phi$-Subproblem}
Collecting~\eqref{eq:objective_surrogate_final_fixed},~\eqref{eq:sinr_compact_phi}, and~\eqref{eq:power_MM_final_convex}, the update of $\boldsymbol\phi$ at iteration $s+1$, denoted by $\boldsymbol\phi^{(s+1)}$ is obtained by solving
\begin{equation}
\begin{aligned}
&\min_{\boldsymbol\phi}~
\boldsymbol\phi^*\mathbf A_y^{(s)}\boldsymbol\phi
+
\Re\left\{
\mathbf b_y^{(s)*}\boldsymbol\phi
\right\}
\\
\mathrm{s.t.}~
&\lambda_p^{(s)}(\boldsymbol\phi^*\boldsymbol\phi)^2
+
\boldsymbol\phi^*(\mathbf K_1+2\sigma_z^2\mathbf I+\mu_p^{(s)}\mathbf I)\boldsymbol\phi
\\
&+2\Re\left\{
\mathbf b_p^{(s)*}\boldsymbol\phi
\right\}
+
\tilde c_p^{(s)}
\le
P_{\max,a},~|\phi_n|\le a_{\max}~( \forall n),\\
&
\boldsymbol\phi^*\mathbf C_k^{(+)}\boldsymbol\phi
+
\Re\{\tilde{\mathbf d}_k^{(s)*}\boldsymbol\phi\}
+
\tilde e_k^{(s)}
\le 0~(\forall k).
\end{aligned}
\label{eq:phi_mm_qcqp_fixed}
\end{equation}
Problem~\eqref{eq:phi_mm_qcqp_fixed} is convex and can be solved efficiently by CVX~\cite{boyd}. After obtaining $\boldsymbol\phi^{(s+1)}$, we update: $\mathbf x^{(s+1)}=\boldsymbol\phi^{(s+1)}\otimes\boldsymbol\phi^{(s+1)},
~
y^{(s+1)}
=
\frac{\mathbf x^{(s+1)*}\mathbf T\mathbf x^{(s+1)}}
{\mathbf x^{(s+1)*}\mathbf U\mathbf x^{(s+1)}}$. The above steps are repeated until convergence. Since the update of $y$ is exact for fixed $\boldsymbol\phi$ obtained in~\eqref{eq:phi_mm_qcqp_fixed}, and vice versa, 
the proposed MM inner iterations generate a nonincreasing sequence of the objective values~\cite{mmtsp, mmtit}, where $\{\boldsymbol\phi^{(s)}\}$ converges to a stationary point of the $\boldsymbol\Phi$-subproblem with fixed $\mathbf W$.

\begin{algorithm}[t]
\caption{Proposed AO Framework}
\label{alg:ao}
\begin{algorithmic}[1]

\State \textbf{Initialize} $\boldsymbol\phi^{(0)}$ and set $r\gets 0$

\While{the AO objective is not converged}

    \Statex \textbf{(1) $\mathbf W$-update for fixed $\boldsymbol\phi^{(r)}$}
    \State Construct $\boldsymbol\Phi^{(r)}=\mathrm{diag}(\boldsymbol\phi^{(r)})$
    \State Solve~\eqref{prbsdp}, obtain $\{\mathbf W_k^\star\}_{k=1}^K$ and recover $\mathbf w_k^{(r+1)}$
    \State Set $\mathbf W^{(r+1)}=[\mathbf w_1^{(r+1)} \cdots \mathbf w_K^{(r+1)}]$

    \Statex \textbf{(2) $\boldsymbol\Phi$-update for fixed $\mathbf W^{(r+1)}$}
    \State $s\gets 0$ and set $\boldsymbol\phi^{(r,0)}\gets \boldsymbol\phi^{(r)}$, $\mathbf x^{(r,0)}$ and $y^{(r,0)}$
    \While{the MM inner loop is not converged}
        \State Solve~\eqref{eq:phi_mm_qcqp_fixed} and obtain $\boldsymbol\phi^{(r,s+1)}$
        \State Update $\mathbf x^{(r,s+1)}$ and $y^{(r,s+1)}$
        \State $s\gets s+1$
    \EndWhile

    \State $\boldsymbol\phi^{(r+1)}\gets \boldsymbol\phi^{(r,s)}$ and $r\gets r+1$

\EndWhile

\State $\mathbf W^\star\leftarrow\mathbf W^{(r)}$ and $\boldsymbol\phi^\star\leftarrow\boldsymbol\phi^{(r)}$
\State \Return $(\mathbf W^\star,\boldsymbol\phi^\star)$
\end{algorithmic}
\end{algorithm}

\subsection{Overall AO Framework and Computational Complexity}
\label{subsec:ao_overall}
According to the above derivations, the proposed joint transmit precoding and ARIS reflection design is summarized in Algorithm~\ref{alg:ao}. Starting from a feasible initialization, the variables are updated in an alternating manner until convergence. At each iteration, the objective value is guaranteed to be non-decreasing, since each subproblem is optimally solved over its feasible set, and the MM-based surrogate function is tight at the current operating point. Consequently, the proposed AO framework produces a monotonically non-decreasing sequence and converges to a stationary point of the original problem.

The computational complexity of the proposed AO framework is determined by the complexity of the two subproblems.
\subsubsection{$\mathbf W$-update}
For fixed $\boldsymbol\phi$,~\eqref{prbsdp} is formulated as a SDP with variables of dimension $N_t \times N_t$ and $K$ linear constraints. Using interior-point methods, the worst-case complexity is dominated by solving a semidefinite cone problem, which scales as $\mathcal{O}\big(\sqrt{K}\max(N_t^4, K^4)\big)$~\cite{sdr}. After solving the SDP, a rank-one solution is recovered via eigenvalue decomposition or Gaussian randomization with complexity $\mathcal{O}(N_t)$ and $\mathcal{O}(N_t^3)$~\cite{sdr}, respectively.

\subsubsection{$\boldsymbol\Phi$-update}
The $\boldsymbol\Phi$-subproblem in~\eqref{eq:phi_mm_qcqp_fixed} is a convex optimization problem with a quadratic objective, $K$ convex quadratic SINR constraints, $N$ amplification gain box constraints, and one additional convex quartic power constraint. By converting the complex variables into their real-valued equivalents, the problem involves $2N$ real variables and $K+N+1$ convex constraints. Hence, the worst-case computational complexity per iteration is on the order of $\mathcal{O}\big(
\sqrt{K+N}N^3
\big)$. 
\subsubsection{Overall complexity}
Let $I_{\text{AO}}$ denote the number of AO iterations and $I_{\text{MM}}$ the number of inner MM iterations for the $\boldsymbol\Phi$-update. Then, the overall complexity of the proposed algorithm is given by
\begin{equation}
\mathcal{O}\big(
I_{\text{AO}}
\big(
I_{\text{MM}}\mathcal{C}_{\boldsymbol\Phi}
+
\mathcal{C}_{W}
\big)
\big),
\end{equation}
where $\mathcal{C}_{W}$ and $\mathcal{C}_{\boldsymbol\Phi}$ denote the complexities of the $\mathbf W$- and $\boldsymbol\Phi$-subproblems, respectively.
\begin{table}[t]
\centering
\caption{System Parameters (Unless Referred)}
\label{tabsim}
\begin{tabular}{l c}
\toprule
\textbf{Parameter} & \textbf{Value} \\
\midrule
Number of RARE-aided users $K$ & 4 \\
Number of BS antenna/RARE elements $(N_t, N_r)$ & (12, 12) \\
Number of user-RARE elements $M$ & 20 \\
Transmit and ARIS power budget $(P_{\max}, P_{\max,a,t})$ & (18, 20)~[dBm]\\
Number of ARIS elements $N$ & 12 \\
SINR thresholds $\gamma_k$ & 8~[dB] \\
Number of radar snapshots $L$ & 1024 \\
Rician $K$-factor $\kappa$ & 5 \\
\bottomrule
\end{tabular}
\end{table}

\section{Simulation Results}
\subsection{Parameter Setup}
Unless otherwise specified, the simulation parameters follow Table~\ref{tabsim}. As illustrated in Fig.~\ref{fig_sim}, BS is located at the origin and simultaneously performs communication and sensing. The communication users are randomly distributed within a circular region centered at $(-10, 26)$~m with a radius of 5~m. A potential target, blocked by obstacles, is located at $(3, 46)$~m, while an ARIS is deployed at $(0, 50)$~m to assist the ISAC operation. The BS-ARIS channel is modeled as Rician fading, and the DoA of the target with respect to the ARIS is set to $\theta=\frac{\pi}{4}$. The path-loss exponents for both the BS-ARIS and ARIS-target links are set to 2.2. For simplicity, we set $\alpha = 1$ and assumed that SI is perfectly cancelled at BS~\cite{fdisac, arisISACtrx}, i.e., $\mathbf H_{\mathrm{SI}} = 0$, since, as can be observed from~\eqref{eq:Er_def} and other relations, $\mathbf H_{\mathrm{SI}}$ only acts as a translation component that does not affect the fundamental structure of the received signal.

For the atomic configuration, the Rydberg energy levels $52D_{5/2}$ and $53P_{3/2}$ are employed to detect the carrier frequency $f=5$~GHz. Based on~\cite{rydpar}, the dipole moment is given by $\boldsymbol{\mu}_{\mathrm{eg}} = [0, 1785.916qa_0, 0]^{\mathrm T}$, where $a_0=5.292\times10^{-11}$~m denotes the Bohr radius and $q=1.602\times10^{-19}$~C is the elementary charge. We set the reference-to-signal ratio (RSR), which measures the power of the reference signal relative to the received communication signal and noise. to a sufficiently high value (15~dB) to satisfy the strong-reference regime required for reliable magnitude-only detection in RARE~\cite{Precoding_atomicMIMO}. The polarization vectors $\{\boldsymbol{\epsilon}_{k,m,b}\}$, $\{\boldsymbol{\epsilon}_{m,n,k,\ell}\}$, as well as those corresponding to $\{\mathbf H_{\mathrm{BU},k}\}$ and $\{\mathbf H_{RB,r}\}$, are independently generated on unit circles orthogonal to their respective incident directions. The results are averaged over $10^3$ Monte Carlo realizations. For the LO-RARE link, due to the negligible separation between the LO-RARE, we approximate $\{\rho_{k,m,b}\}$ by a constant $\rho_b$, while $\{\phi_{k,m,b}\}$ are modeled as independent and uniformly distributed over $[0,2\pi)$. The same assumption is applied to the corresponding parameters of $\{\mathbf b_r[\ell]\}$.


\subsection{Reliability and Performance Results}

\begin{figure}[t]
    \centering
    \subfloat[]{%
        \includegraphics[width=0.13\textwidth]{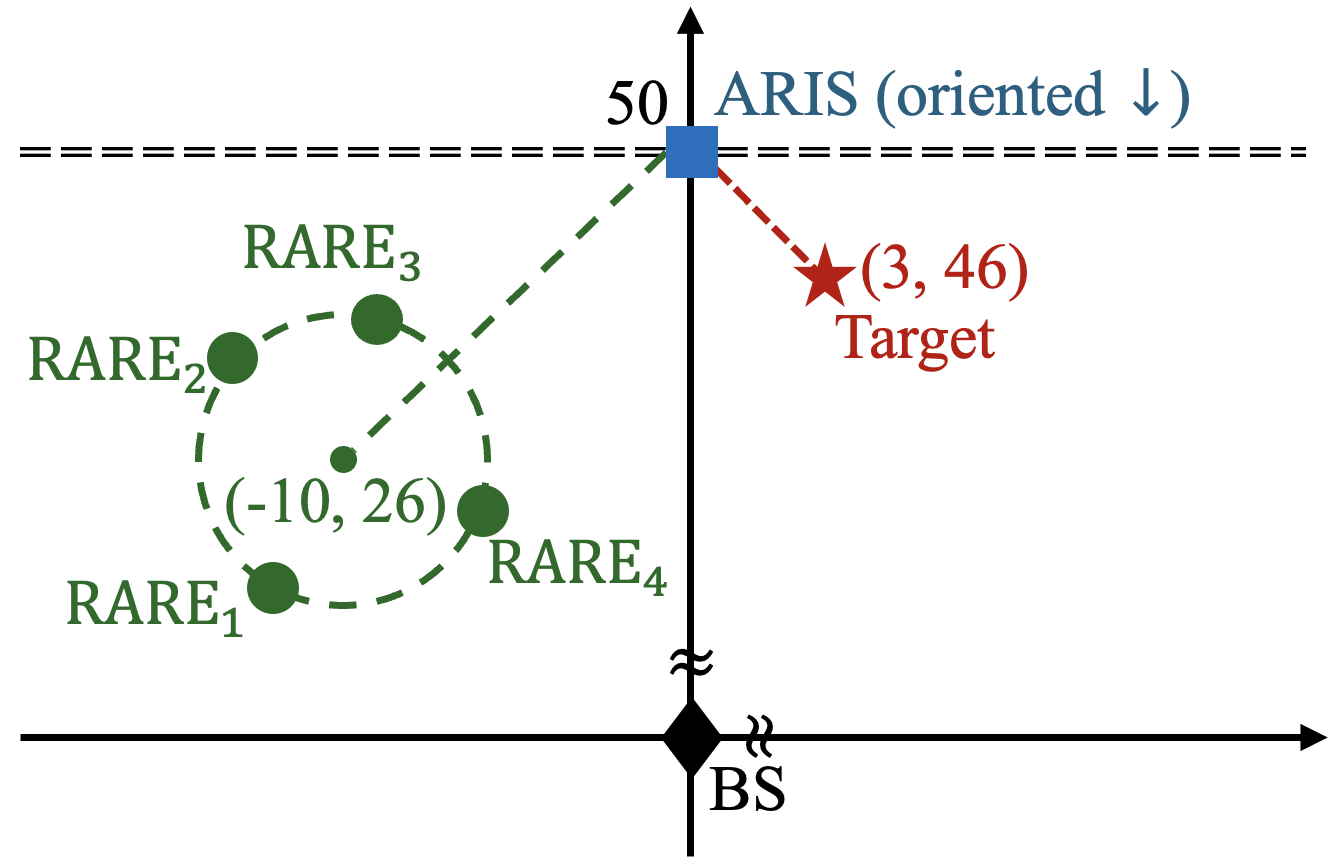}
        \label{fig_sim}%
    }
    \subfloat[]{%
        \includegraphics[width=0.12\textwidth]{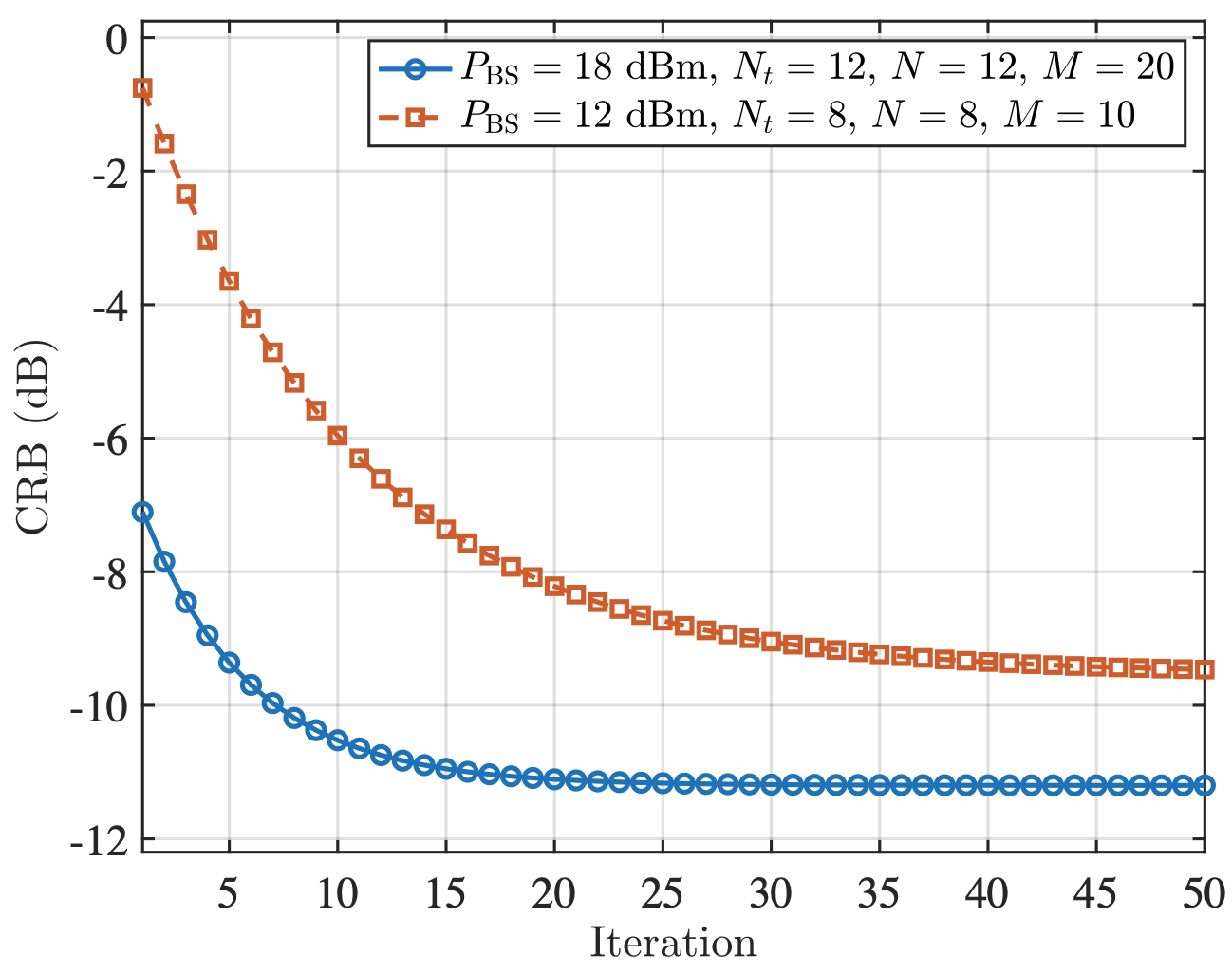}
        \label{fig_conv}%
    }
    \caption{(a) Illustration of simulation environment and (b) convergence of the proposed AO framework.}
    \label{fig_0}
\end{figure}

The convergence behavior of the proposed AO framework is depicted in Fig.~\ref{fig_conv}, where two distinct system parameter configurations are considered for comparison. The results demonstrate that the proposed framework attains convergence within a limited number of iterations and exhibits fast and stable convergence performance.

\begin{figure}[t]
    \centering
    \subfloat[]{%
        \includegraphics[width=0.115\textwidth]{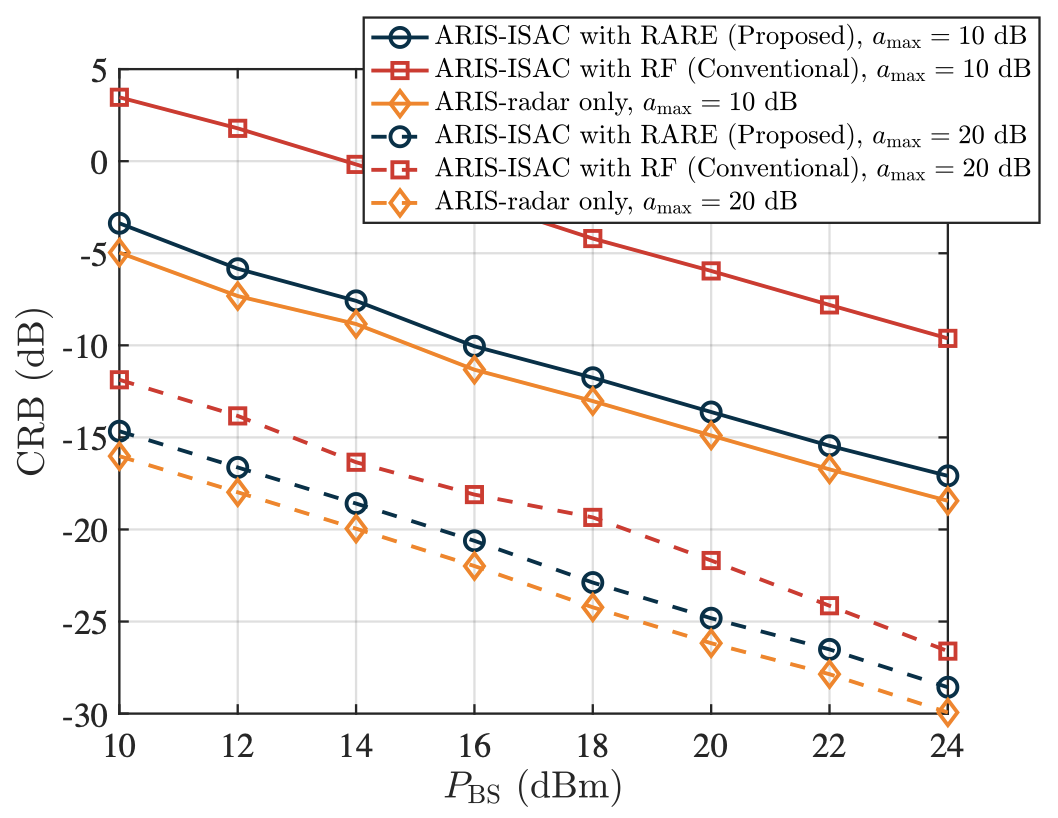}
        \label{fig_pbs}%
    }
    \subfloat[]{%
        \includegraphics[width=0.115\textwidth]{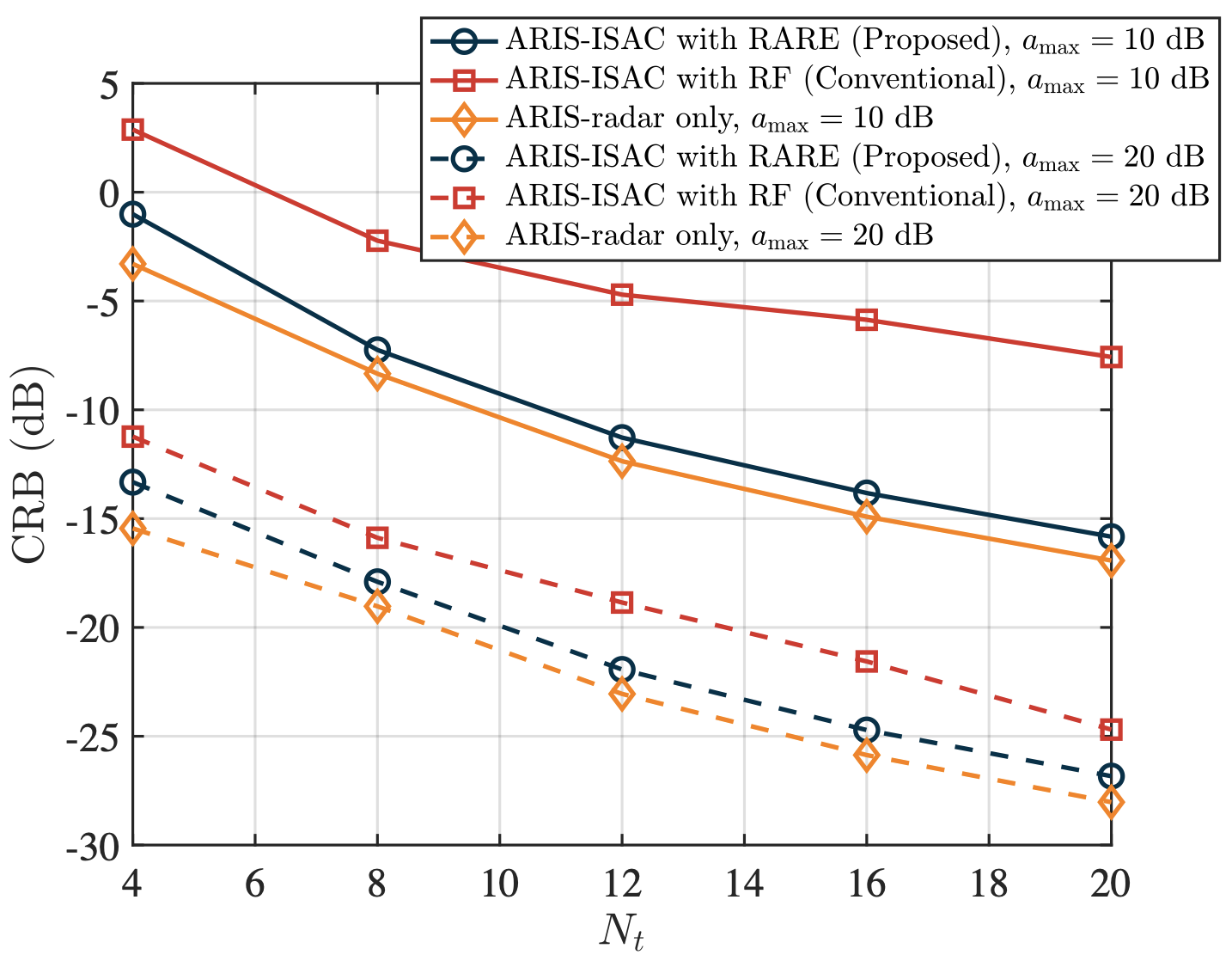}
        \label{fig_nt}%
    }
        \subfloat[]{%
        \includegraphics[width=0.115\textwidth]{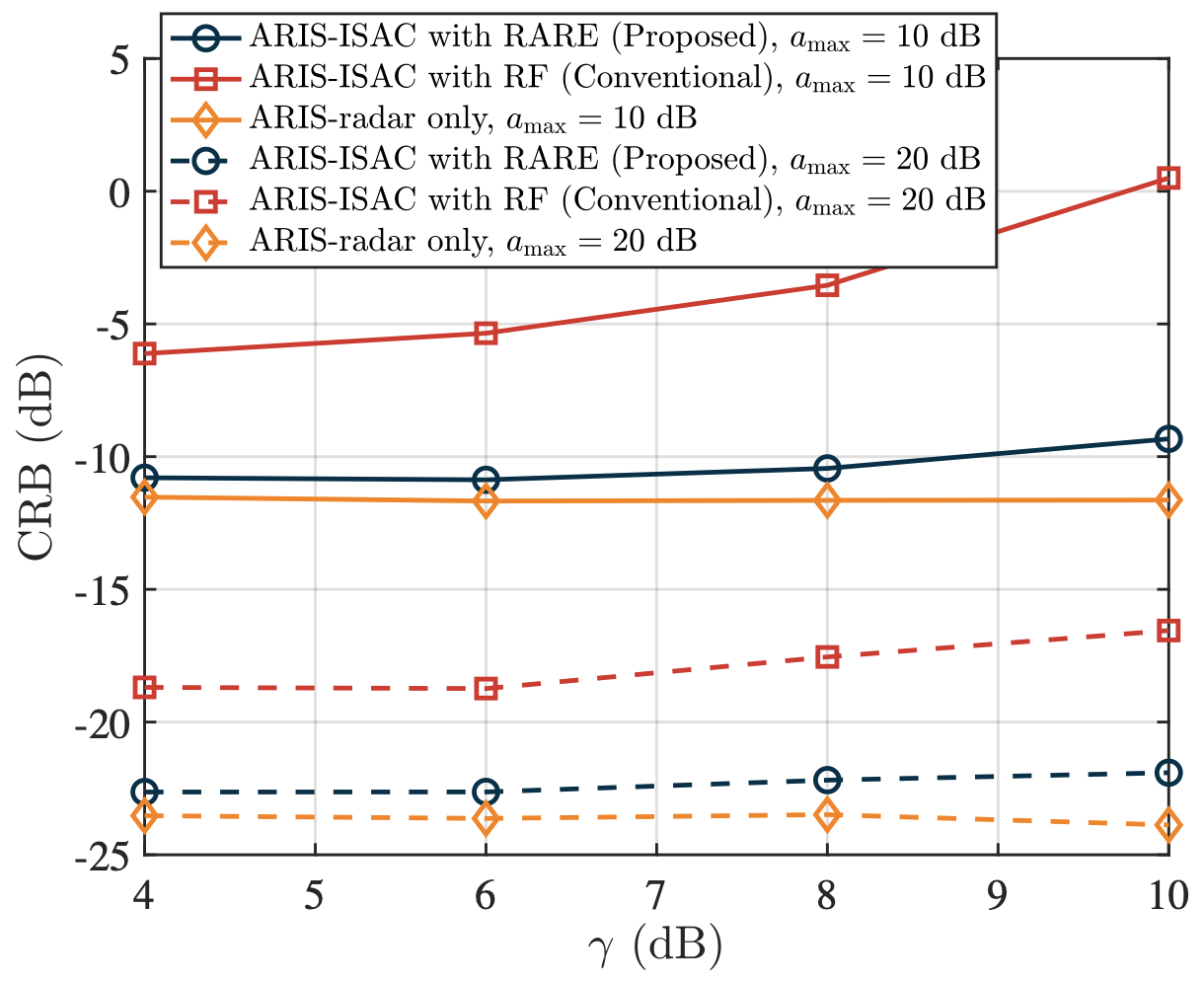}
        \label{fig_sinr}%
    }
    \subfloat[]{%
        \includegraphics[width=0.115\textwidth]{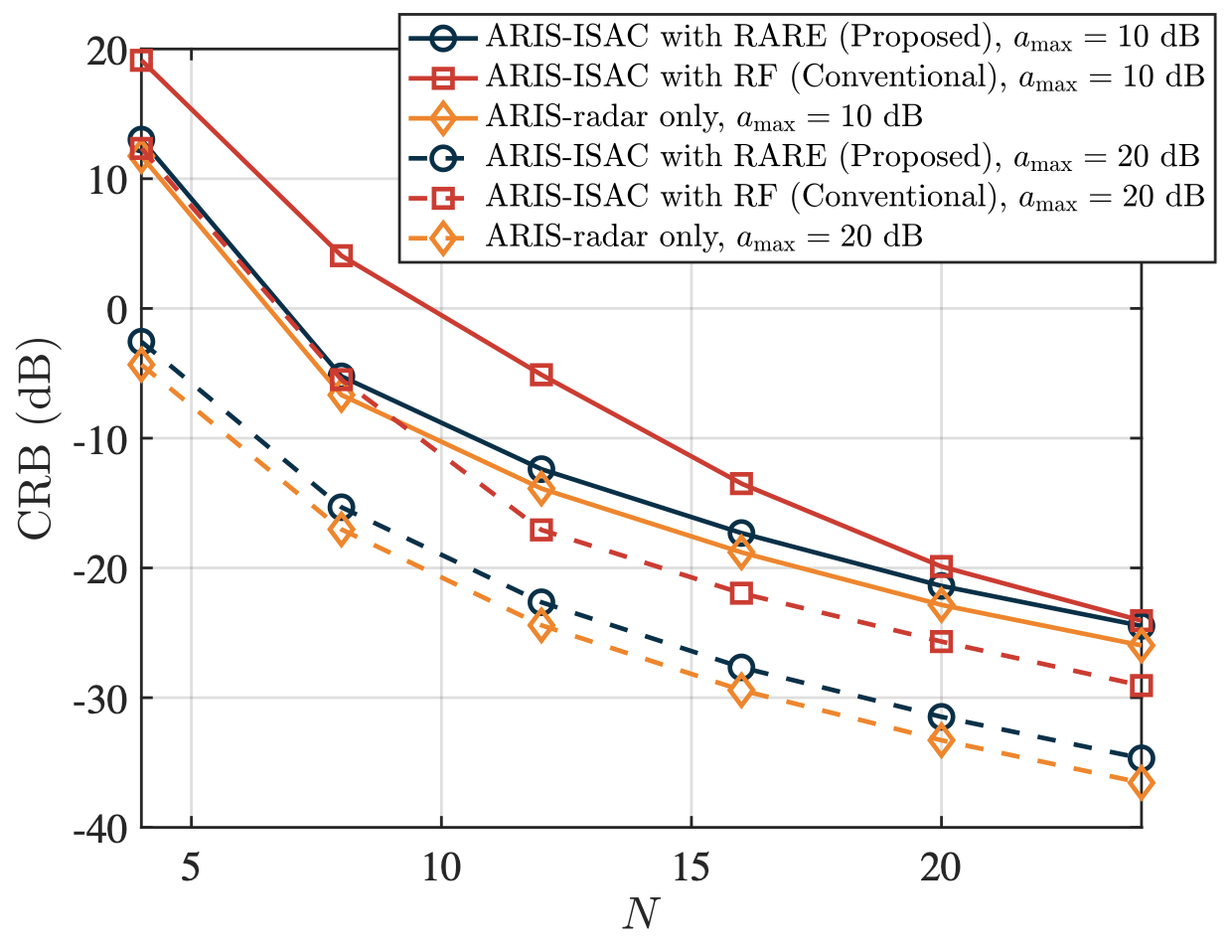}
        \label{fig_nn}%
    }
    \caption{CRB versus (a) $P_{\mathrm{BS}}$, (b) $N_t$, (c) $\gamma_k$, and (d) $N$ for the proposed ARIS-ISAC with RARE and baselines.}
    \label{fig_1}
\end{figure}

Fig.~\ref{fig_pbs} presents the CRB versus $P_{\mathrm{BS}}$ under different system configurations. As expected, the CRB decreases monotonically with increasing $P_{\mathrm{BS}}$ for all frameworks, since higher transmit power improves the received echo signal quality. In particular, the proposed RARE-enabled ARIS-ISAC framework consistently maintains a clear performance advantage over the conventional RF-based counterpart under considered settings, demonstrating its effectiveness in exploiting the given propagation environment. This improvement stems from the synergy between the RARE-based sensing and communications architecture with lower QSN than JNTN and the ARIS, where the amplified cascaded links enhance the useful echo signal. Moreover, the proposed framework closely approaches the performance of the radar-only scheme, indicating that it effectively preserves sensing capability even in the presence of communication functionality. This highlights that the proposed design successfully balances the sensing-communication trade-off while maintaining near-optimal sensing performance. Besides, the case of $a_{\max} = 20$~dB is always superior to the case of $a_{\max} = 10$~dB, which means that a wider range of amplitude variation can bring a higher DoF and lead to better performance.

Fig.~\ref{fig_nt} illustrates the CRB versus $N_t$, where the numbers of antennas and RARE elements at BS are simultaneously increased. Similar to the previous observations, the proposed RARE-enabled ARIS-ISAC framework consistently outperforms the conventional RF-based counterpart and achieves performance close to the ARIS-assisted radar-only scheme. The performance improvement with increasing $N_t$ is attributed to the enhanced spatial diversity and array gain provided by the enlarged RF and RARE arrays at BS with lower QSN compared to JNTN, where increasing $a_{\max}$ also improves the performance, showing that a larger amplification range more effectively leverages the expanded array dimensions.

Fig.~\ref{fig_sinr} displays the CRB versus $\forall\gamma_k=\gamma$. As $\gamma$ increases, the CRB of all ISAC schemes gradually increases, since more resources are allocated to satisfy the communication QoS, thereby degrading sensing performance. Compared to the ARIS-assisted radar-only scheme, the ISAC frameworks incur a performance loss that becomes more pronounced at higher $\gamma$, revealing the inherent sensing-communication trade-off. Nevertheless, the proposed RARE-enabled ARIS-ISAC framework shows much stronger robustness than the conventional RF-based counterpart, maintaining a more marginal CRB degradation across the entire $\gamma$ range; this implies that sufficient power margin is available in proposed framework to satisfy the communication requirement without significantly sacrificing sensing performance. This is attributed to the low-QSN RARE architecture combined with ARIS, which effectively preserves sensing capability under communication constraints. As a result, the proposed scheme remains close to the radar-only performance, especially in the low-$\gamma$ regime. Furthermore, increasing $a_{\max}$ consistently improves the CRB and alleviates the degradation, highlighting the benefit of enhanced active amplification.

Fig.~\ref{fig_nn} illustrates the CRB versus $N$. As expected, the CRB decreases monotonically with increasing $N$ for all schemes, since a larger ARIS provides more exploitable spatial DoF and enhances the cascaded channel gain. However, the improvement gradually saturates with increasing $N$, since from~\eqref{eq:P_constraint}, it reduces $P_{\max,a}$ as more ARIS elements are added. We can observe that the proposed RARE-enabled ARIS-ISAC framework significantly outperforms the conventional RF-based counterpart and achieves performance close to the ARIS-assisted radar-only scheme. As expected, the CRB of all scenarios decreases with increasing $N$ due to the higher exploitable spatial DoF provided by the enlarged RIS, while the integration of the low-QSN RARE architecture further enhances the overall sensing performance.

\begin{figure}[t]
    \centering
    \subfloat[]{%
        \includegraphics[width=0.13\textwidth]{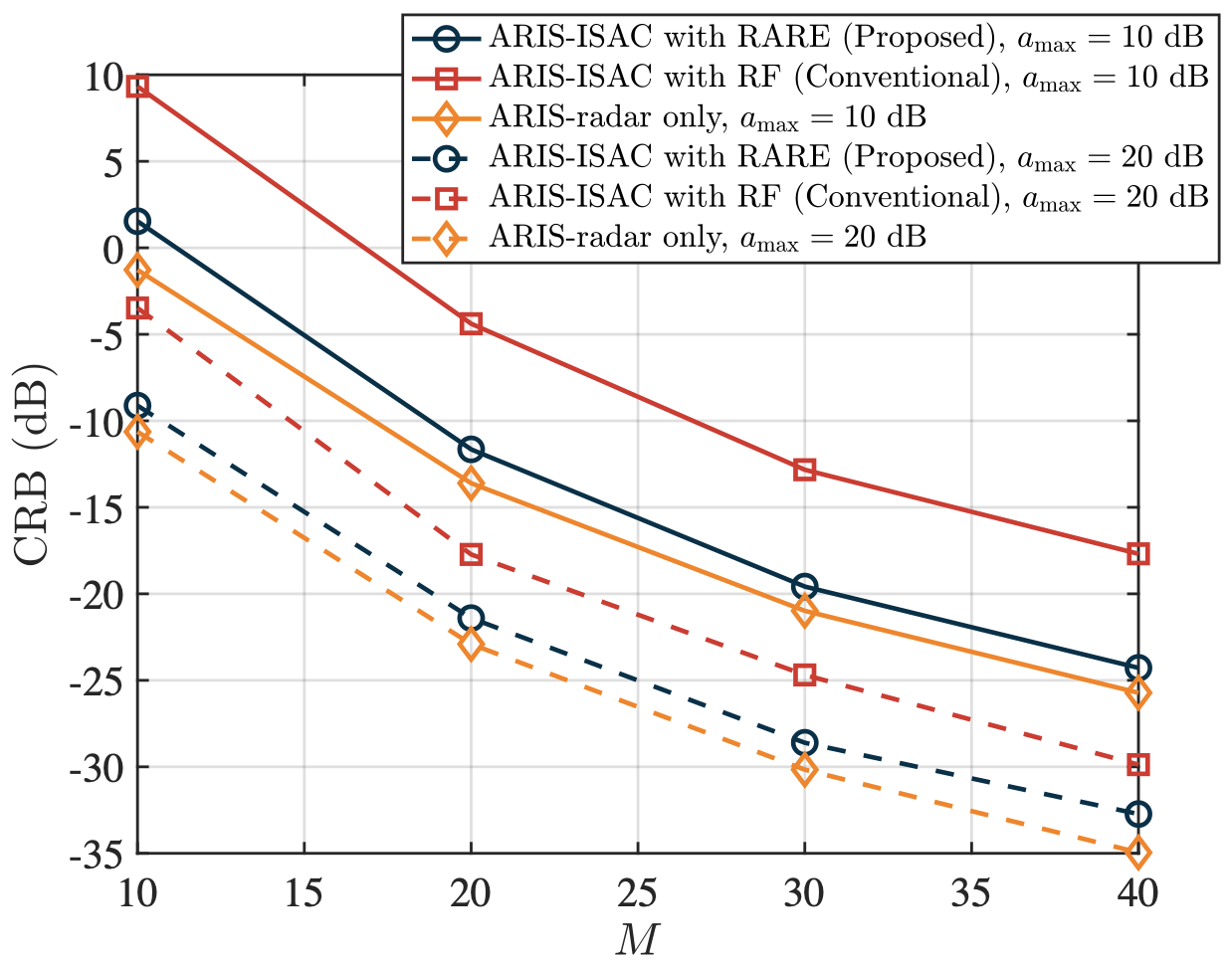}
        \label{fig_mm}%
    }
    \subfloat[]{%
        \includegraphics[width=0.13\textwidth]{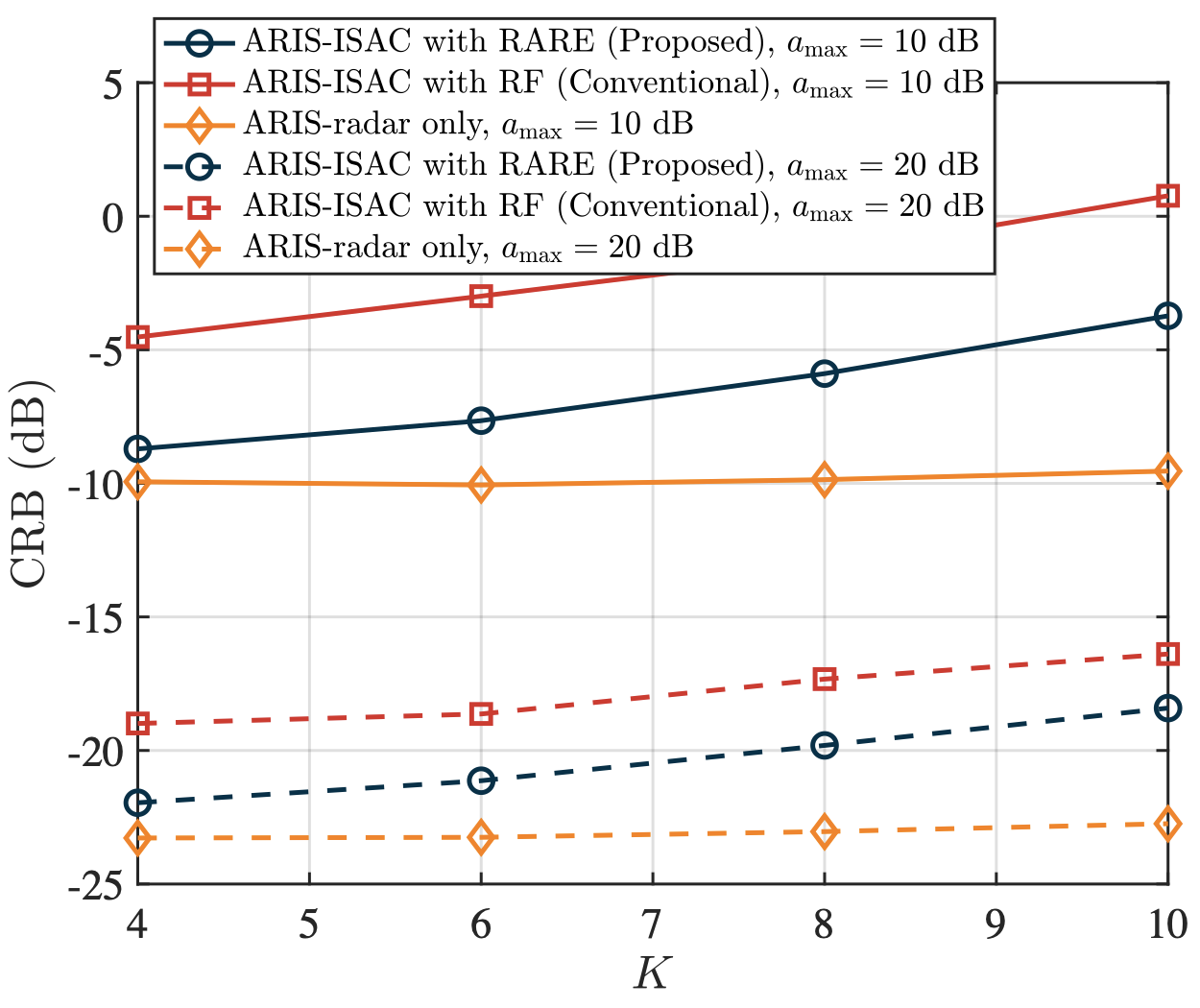}
        \label{fig_kk}%
    }
        \subfloat[]{%
        \includegraphics[width=0.13\textwidth]{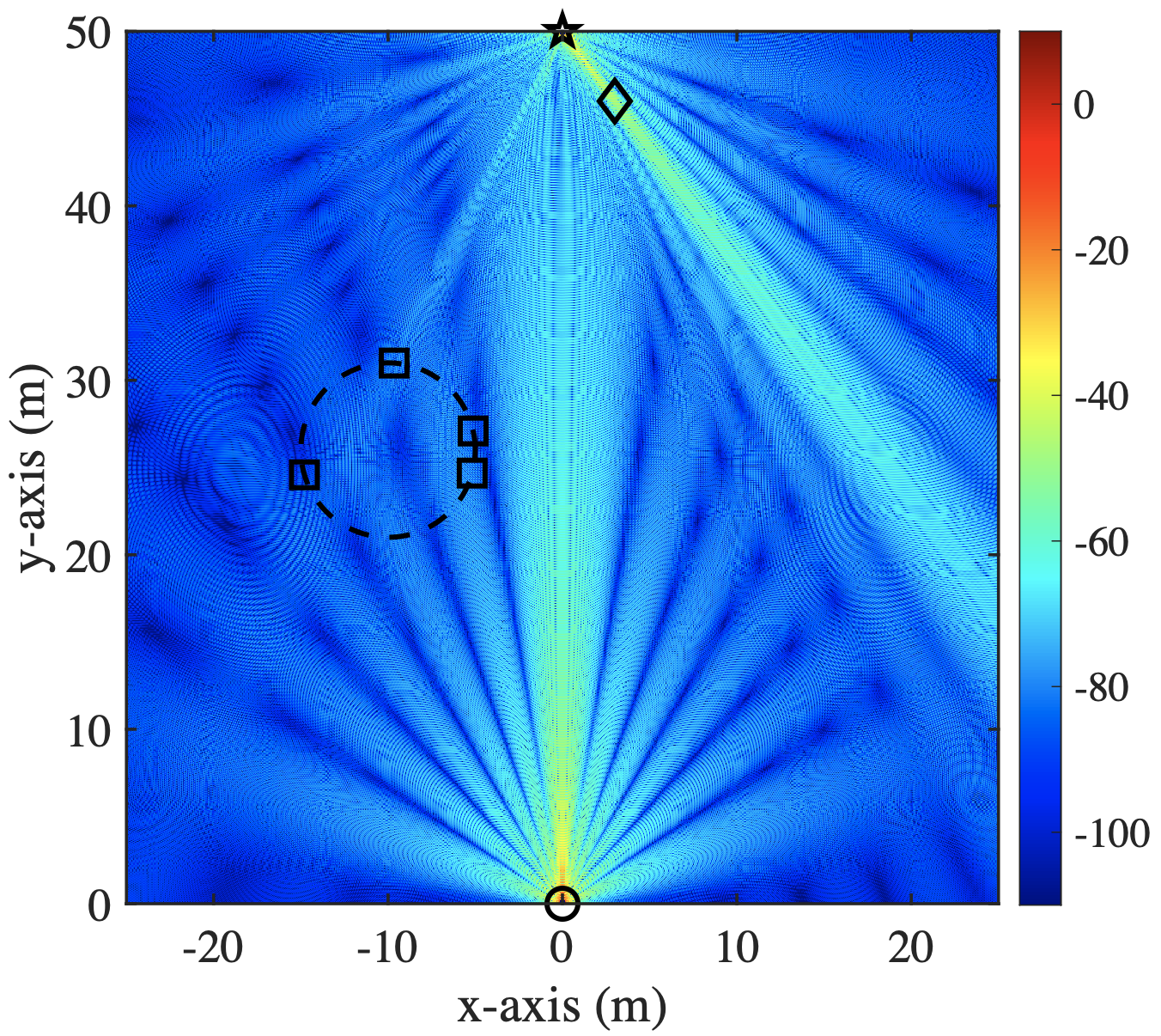}
        \label{fig_beam}%
    }
    \caption{CRB versus (a) $M$ and (b) $K$ for the proposed ARIS-ISAC with RARE and baselines, and (c) beampattern of the ARIS-ISAC system with RARE with $N=24$, and $a_{\max}=10$~dB (BS: circle on origin; ARIS: star; target: diamond; users: square).}
    \label{fig_2}
\end{figure}

Fig.~\ref{fig_mm} depicts the CRB versus $M$. As expected, the CRB decreases monotonically with increasing $M$ for all schemes, since a larger RARE array at Rx enhances the effective signal strength and provides additional spatial DoF; increasing $a_{\max}$ also consistently improves the CRB, indicating that a larger amplification more effectively leverages the expanded RARE array at Rx. We can observe that the proposed RARE-enabled ARIS-ISAC framework significantly outperforms the conventional RF-based counterpart and achieves performance close to the ARIS-assisted radar-only scheme. This gain is attributed to the enlarged communication-side RARE array, which, despite being designed for communication, improves the overall signal quality through its inherently lower QSN. When combined with ARIS-based amplification, this enhanced signal propagation indirectly benefits the sensing process by strengthening the cascaded echo signal, thereby improving DoA estimation accuracy. 

Fig.~\ref{fig_kk} demonstrates the CRB as a function of $K$. As expected, the CRB increases with growing $K$, since more resources are allocated to the communication function to satisfy the SINR requirements, leading to degraded sensing performance. Nevertheless, the proposed RARE-enabled ARIS-ISAC framework consistently outperforms the conventional RF-based counterpart and remains close to the ARIS-assisted radar-only scheme, in both large and small $a_{\max}$, demonstrating its robustness against increasing communication load. This is mainly attributed to the low-QSN RARE architecture, which effectively alleviates the sensing degradation under stringent communication demands.

In Fig.~\ref{fig_beam}, we depict the beampattern of the ARIS-ISAC system with RARE based on the real-part effective channel in~\eqref{eq:stacked_model}. The BS forms a directive beam toward the ARIS to efficiently deliver energy, while the ARIS reshapes the incident signals to generate reflection beams toward both the target and the user region. In particular, a sharp beam is steered toward the target due to coherent combining for CRB minimization, whereas a wider beam covers the user cluster to ensure reliable communication, thereby validating the effectiveness of the proposed framework.

\section{Conclusion}
In this paper, we investigated a RARE-enabled ARIS-assisted ISAC framework and developed a unified design that jointly optimized the BS beamforming and ARIS reflection coefficients under practical communication and sensing constraints. By explicitly incorporating the real-domain observation structure of RARE and the amplified-noise characteristics of ARIS, we established a CRB-driven optimization framework with an efficient AO-based solution. Numerical results demonstrated that the proposed RARE-aware design significantly outperformed conventional RF-based ISAC schemes and achieved performance close to the radar-only benchmark, while effectively balancing the sensing-communication trade-off. These findings highlighted the benefit of combining low-QSN RARE with ARIS-based amplification in mitigating multiplicative fading and preserving sensing accuracy, revealing the strong potential of RARE for quantum-enhanced ISAC.

\appendices
\section{Derivation of the Lifted Representation in~\eqref{eq:utv_x_form}}
\label{app:lifting_utv}

\subsection{Vectorized Representation of $\mathbf Q(\boldsymbol\Phi,\theta)$ and $\dot{\mathbf Q}(\boldsymbol\Phi,\theta)$}
Using the identity $\mathrm{vec}(\mathbf A\mathbf X\mathbf B)
=
(\mathbf B^{\mathrm T}\otimes \mathbf A)\mathrm{vec}(\mathbf X)$, we obtain from~\eqref{eq:Q_phi_exact} that
\begin{equation}
\mathrm{vec}\left(\mathbf Q(\boldsymbol\Phi,\theta)\right)
=
(\mathbf B_0^{\mathrm T}\otimes \tilde{\mathbf B}_0^{\mathrm T})
\mathrm{vec}(\boldsymbol\phi\boldsymbol\phi^{\mathrm T})
=
\mathbf G_0\mathbf x,
\label{eq:app_vecQ}
\end{equation}
where $\mathbf G_0 \triangleq \mathbf B_0^{\mathrm T}\otimes \tilde{\mathbf B}_0^{\mathrm T}$. Likewise, from~\eqref{eq:Qdot_phi_exact},
\begin{equation}
\begin{aligned}
\mathrm{vec}\left(\dot{\mathbf Q}(\boldsymbol\Phi,\theta)\right)&=
(\mathbf B_0^{\mathrm T}\otimes \tilde{\mathbf B}_1^{\mathrm T})
\mathrm{vec}(\boldsymbol\phi\boldsymbol\phi^{\mathrm T})+
(\mathbf B_1^{\mathrm T}\otimes \tilde{\mathbf B}_0^{\mathrm T})
\mathrm{vec}(\boldsymbol\phi\boldsymbol\phi^{\mathrm T})\\
&=\mathbf G_1\mathbf x,
\end{aligned}
\label{eq:app_vecQdot}
\end{equation}
where $\mathbf G_1
\triangleq
\mathbf B_0^{\mathrm T}\otimes \tilde{\mathbf B}_1^{\mathrm T}
+
\mathbf B_1^{\mathrm T}\otimes \tilde{\mathbf B}_0^{\mathrm T}$. 
\subsection{Derivation of $u(\boldsymbol\phi)$} 

From~\eqref{utvtrace}, we have $u(\boldsymbol\phi)
=
L\mathrm{tr}
\left\{
\mathbf Q
\mathbf R_W
\mathbf Q^{*}
\mathbf M_r
\right\}$. Applying the identity $\mathrm{tr}(\mathbf A\mathbf B\mathbf C\mathbf D)
=
\mathrm{vec}(\mathbf A)^{*}
(\mathbf D^{\mathrm T}\otimes \mathbf B)
\mathrm{vec}(\mathbf C)$ with $\mathbf A=\mathbf Q,
\mathbf B=\mathbf R_W,
\mathbf C=\mathbf Q,
\mathbf D=\mathbf M_r$, we obtain
\begin{equation}
u(\boldsymbol\phi)
=
L\mathrm{vec}(\mathbf Q)^{*}
(\mathbf M_r^{\mathrm T}\otimes \mathbf R_W)
\mathrm{vec}(\mathbf Q).
\label{eq:app_u_vec1}
\end{equation}
Substituting~\eqref{eq:app_vecQ} into~\eqref{eq:app_u_vec1} yields
\begin{equation}
\begin{aligned}
u(\boldsymbol\phi)
&=
L(\mathbf G_0\mathbf x)^{*}
(\mathbf M_r^{\mathrm T}\otimes \mathbf R_W)
(\mathbf G_0\mathbf x)\\
&=
\mathbf x^{*}
\Big(
L\mathbf G_0^{*}
(\mathbf M_r^{\mathrm T}\otimes \mathbf R_W)
\mathbf G_0
\Big)\mathbf x.
\end{aligned}
\label{eq:app_u_vec2}
\end{equation}
Hence $u(\boldsymbol\phi)=\mathbf x^{*}\mathbf U\mathbf x,
~
\mathbf U
=
L\mathbf G_0^{*}
(\mathbf M_r^{\mathrm T}\otimes \mathbf R_W)
\mathbf G_0$.
Furthermore, since $\mathbf R_w$ is real and symmetric with $\mathbf R_w \succ \mathbf 0$, so is $\mathbf R_w^{-1}$. Since $\mathbf D_{b,r}$ is a diagonal unitary matrix: $\mathbf M_r
=
\mathbf D_{b,r}^{*}\mathbf R_w^{-1}\mathbf D_{b,r}
\succ \mathbf 0$, which follows from unitary similarity. Then, $\mathbf M_r^{\mathrm T}$ can be written as
\begin{equation}
\label{mdsim}
\mathbf M_r^{\mathrm T}
=
(\mathbf D_{b,r}^{*}\mathbf R_w^{-1}\mathbf D_{b,r})^{\mathrm T}
=
\mathbf D_{b,r}^{\mathrm T}\mathbf R_w^{-1}\bar{\mathbf D}_{b,r}\overbrace{=}^{(a)}\mathbf D_{b,r}\mathbf R_w^{-1}\mathbf D_{b,r}^{*},
\end{equation}
where (a) holds since $\mathbf D_{b,r}$ is diagonal with unit-modulus entries, we have $\mathbf D_{b,r}^{\mathrm T}=\mathbf D_{b,r}$ and $\bar{\mathbf D}_{b,r}=\mathbf D_{b,r}^{*}$. Hence, $\mathbf M_r^{\mathrm T}$ is also obtained via a unitary similarity transformation of $\mathbf R_w^{-1}$, implying $\mathbf M_r^{\mathrm T}\succ \mathbf 0$. Therefore, combining with $\mathbf R_W\succ 0$, by the property of the Kronecker product: $\mathbf M_r^{\mathrm T}\otimes \mathbf R_W \succ \mathbf 0$, and by applying the congruence transformation with $\mathbf G_0$, we obtain $\mathbf U
=
L\mathbf G_0^{*}
(\mathbf M_r^{\mathrm T}\otimes \mathbf R_W)
\mathbf G_0
\succ \mathbf 0$.
\subsection{Derivation of $t(\boldsymbol\phi)$}
From~\eqref{utvtrace}, we have $t(\boldsymbol\phi)=L\mathrm{tr}\left\{\dot{\mathbf Q}\mathbf R_W\mathbf Q^{*}\mathbf M_r\right\}$. Applying $\mathrm{tr}(\mathbf A\mathbf B\mathbf C\mathbf D)
=
\mathrm{vec}(\mathbf A)^{*}
(\mathbf D^{\mathrm T}\otimes \mathbf B)
\mathrm{vec}(\mathbf C)$ again with $
\mathbf A=\dot{\mathbf Q},
\mathbf B=\mathbf R_W,
\mathbf C=\mathbf Q,
\mathbf D=\mathbf M_r$, we obtain
\begin{equation}
t(\boldsymbol\phi)
=
L\mathrm{vec}(\dot{\mathbf Q})^{*}
(\mathbf M_r^{\mathrm T}\otimes \mathbf R_W)
\mathrm{vec}(\mathbf Q).
\label{eq:app_t_vec1}
\end{equation}
Substituting~\eqref{eq:app_vecQ} and~\eqref{eq:app_vecQdot} gives
\begin{equation}
\begin{aligned}
t(\boldsymbol\phi)
&=
L(\mathbf G_1\mathbf x)^{*}
(\mathbf M_r^{\mathrm T}\otimes \mathbf R_W)
(\mathbf G_0\mathbf x)\\
&=
\mathbf x^{*}
\Big(
L\mathbf G_1^{*}
(\mathbf M_r^{\mathrm T}\otimes \mathbf R_W)
\mathbf G_0
\Big)\mathbf x.
\end{aligned}
\label{eq:app_t_vec2}
\end{equation}
Therefore $t(\boldsymbol\phi)=\mathbf x^{*}\mathbf T\mathbf x,
~
\mathbf T
=
L\mathbf G_1^{*}
(\mathbf M_r^{\mathrm T}\otimes \mathbf R_W)
\mathbf G_0$.
\subsection{Derivation of $v(\boldsymbol\phi)$}

From~\eqref{utvtrace}, we have $v(\boldsymbol\phi)
=
L\mathrm{tr}
\left\{
\dot{\mathbf Q}
\mathbf R_W
\dot{\mathbf Q}^{*}
\mathbf M_r
\right\}$. Applying $\mathrm{tr}(\mathbf A\mathbf B\mathbf C\mathbf D)
=
\mathrm{vec}(\mathbf A)^{*}
(\mathbf D^{\mathrm T}\otimes \mathbf B)
\mathrm{vec}(\mathbf C)$ again with $\mathbf A=\dot{\mathbf Q},
\mathbf B=\mathbf R_W,
\mathbf C=\dot{\mathbf Q},
\mathbf D=\mathbf M_r$, we obtain
\begin{equation}
v(\boldsymbol\phi)
=
L\mathrm{vec}(\dot{\mathbf Q})^{*}
(\mathbf M_r^{\mathrm T}\otimes \mathbf R_W)
\mathrm{vec}(\dot{\mathbf Q}).
\label{eq:app_v_vec1}
\end{equation}
Substituting~\eqref{eq:app_vecQdot} into~\eqref{eq:app_v_vec1} yields
\begin{equation}
\begin{aligned}
v(\boldsymbol\phi)
&=
L(\mathbf G_1\mathbf x)^{*}
(\mathbf M_r^{\mathrm T}\otimes \mathbf R_W)
(\mathbf G_1\mathbf x)\\
&=
\mathbf x^{*}
\Big(
L\mathbf G_1^{*}
(\mathbf M_r^{\mathrm T}\otimes \mathbf R_W)
\mathbf G_1
\Big)\mathbf x.
\end{aligned}
\label{eq:app_v_vec2}
\end{equation}
Thus $v(\boldsymbol\phi)=\mathbf x^{*}\mathbf V\mathbf x,
~
\mathbf V
=
L\mathbf G_1^{*}
(\mathbf M_r^{\mathrm T}\otimes \mathbf R_W)
\mathbf G_1$. The positive definiteness of $\mathbf V$ can be similarly proven with $\mathbf U$, which completes the derivation of~\eqref{eq:utv_x_form}.
\hfill$\blacksquare$
\bibliographystyle{IEEEtran}
\bibliography{IEEEexample}

\end{document}